\documentclass[aps,amsmath,amssymb,twocolumn,showpacs,nofootinbib]{revtex4}

\usepackage{graphicx,color}
 \graphicspath{{./}{./}}
\usepackage{dcolumn}
\usepackage{bm}
\usepackage{feynmp-auto}

\definecolor{labelkey}{cmyk}{.4,.2,0,0}

\newcommand{\be}{\begin{equation}}
\newcommand{\ee}{\end{equation}}
\newcommand{\bea}{\begin{eqnarray}}
\newcommand{\eea}{\end{eqnarray}}

\newcommand{\nn}{\nonumber }

\newcommand{\Z}{\mathbb{Z}}

\begin{document}

\title{Universal correlations between shocks in the ground state of elastic interfaces in disordered media}
\author{Thimoth\'ee Thiery, Pierre Le Doussal and Kay J\"org Wiese} 
\affiliation{CNRS-Laboratoire
de Physique Th{\'e}orique de l'Ecole Normale Sup{\'e}rieure, 
PSL Research University, Sorbonne Universit\'es, UPMC,
24 rue
Lhomond,75005 Paris, France.}

\begin{abstract}
The ground state of an elastic interface in a disordered medium undergoes 
collective jumps upon variation of external parameters. These mesoscopic jumps are called shocks, or
static avalanches. Submitting the interface to a parabolic potential centered at $w$, 
we study the avalanches which occur as $w$ is varied. We are interested in the  correlations
between the avalanche sizes $S_1$ and $S_2$   occurring   
at positions $w_1$ and $w_2$.
Using the Functional Renormalization Group (FRG), we show that    correlations exist  for realistic interface models below their upper critical dimension. 
Notably,  the 
connected moment
$ \langle S_1 S_2 \rangle^c$ is up to a prefactor {\em exactly}  the  renormalized disorder correlator, itself a function of  $|w_2-w_1|$. The latter is the universal function at the center of the FRG; hence correlations between shocks are universal as well. All moments and the full joint probability distribution are computed to first non-trivial order in an $\epsilon$-expansion below the upper critical dimension. To quantify the local nature of the coupling between avalanches, we calculate the correlations
of their  local  jumps. We finally test our predictions against simulations of a particle in random-bond and  random-force disorder, with surprisingly good agreement.
\end{abstract}
\maketitle

\section{Introduction}

The model of an elastic interface in a disordered medium has been put forward as a relevant description for a large number of systems \cite{DSFisher1998,BlatterFeigelmanGeshkenbeinLarkinVinokur1994,NattermannScheidl2000,GiamarchiLeDoussalBookYoung}. Examples include domain walls in soft magnets \cite{ZapperiCizeauDurinStanley1998,DurinZapperi2000}, fluid contact lines on a rough surface \cite{MoulinetGuthmannRolley2002,LeDoussalWieseMoulinetRolley2009}, strike-slip faults in geophysics \cite{BenZionRice1993,FisherDahmenRamanathanBenZion1997}, fracture in brittle materials \cite{Ponson2008,SantucciGrobToussaint2010,BonamySantucciPonson2008} or imbibition fronts \cite{PlanetRamonSantucci2009}. An important common property of these systems is that their response to an applied field is not smooth but rather proceeds via jumps extending over a broad range of space and time scales. As a consequence,  understanding the properties and the universality of avalanche processes has received a lot of attention in the past  years \cite{SethnaDahmenMyers2001,AragonKoltonDoussalWieseJagla2016,DurinBohnCorreaSommerDoussalWiese2016}. 

A problem of outstanding interest is to quantify the correlations between successive avalanches. In the context of earthquakes those are linked to the notion of {\it aftershocks}, whose statistics is characterized through phenomenological laws such as the Omori law \cite{Omori1894}.\ Several  mechanisms
have been advanced to explain these strong correlations, all involving an additional dynamical variable  \cite{BurridgeKnopoff1967,JaglaKolton2009}. For elastic interfaces, correlations between avalanches were yet only studied as a result  of such  additional degrees of freedom in the interface dynamics,   as relaxation processes \cite{JaglaLandesRosso2014,Jagla2014} or memory effects \cite{DobrinevskiLeDoussalWiese2013}. In this work,  we show that even in the absence of such mechanisms, {\em\ avalanches in elastic interfaces  are generically correlated below their upper critical dimension}. These correlations are universal. 

Let us   emphasize that the goal of this paper is {\it not} to understand or explain the aftershock statistics observed in earthquakes, for which additional mechanisms such as those discussed above are necessary. Rather, it is to emphasize that for disordered elastic systems, {\em   except for mean-field models, correlations between avalanches always exist}.  
A precise quantitative understanding of these correlations is necessary  to correctly quantify  correlations induced by additional mechanisms. In systems where the description by the standard elastic-interface model is accurate (without additional mechanisms) our results quantify   the correlations between avalanches. To our knowledge, these correlations have up to now been ignored in theoretical or experimental work. It would thus be interesting to quantify them better, in order to access  universality, or lack thereof, in various avalanche processes.

In this article we study the correlations between the sizes and locations of shocks in the ground state (also called ``static avalanches'') of elastic interfaces in disordered media. These static avalanches are close cousins of the (dynamic)\ avalanches observed in the interface dynamics at depinning. As we discuss below, we expect most of our results to hold for both classes. Our study is conducted using the Functional Renormalization Group (FRG). Originally introduced as a powerful tool to study the universal properties of the statics and dynamics (at the depinning transition) of elastic interfaces in disordered media \cite{DSFisher1986,NattermannStepanowTangLeschhorn1992,NarayanDSFisher1992b,NarayanDSFisher1993a,ChauveLeDoussalWiese2000a,LeDoussalWieseChauve2002,LeDoussalWieseChauve2003}, the   FRG has been recently adapted to the study of avalanches \cite{LeDoussalWiese2008c,LeDoussalWiese2011b,LeDoussalWiese2012a,DelormeLeDoussalWiese2016,ThieryLeDoussalWiese2015,DobrinevskiLeDoussalWiese2014a}. It has notably led to a rigorous identification  of the relevant mean-field theory for the statistics of single avalanches: the Brownian-Force Model (BFM), a multidimensional generalization of the celebrated {Alessandro-Beatrice-Bertotti-Montorsi (ABBM) model \cite{AlessandroBeatriceBertottiMontorsi1990,AlessandroBeatriceBertottiMontorsi1990b}. Interestingly, the FRG  allows to go beyond mean-field theory and to compute in a controlled way avalanche observables in an expansion in $\epsilon= d_{\rm uc}-d$ where $d$ is the interface dimension, and $d_{\rm uc}$  the upper critical dimension of the problem. The latter depends on the range of the elastic interactions, with $d_{\rm uc}=4$ for short-ranged (SR) elasticity and $d_{\rm uc}=2$ for the usual long-ranged (LR) elasticity.

\medskip

The outline of this article is as follows: In section \ref{s:main-results} we summarize our results, preceded by a definition of the relevant observables. 
In  Section \ref{Sec:Model} we introduce the model and the observables we are interested in.  Section \ref{Sec:Global} contains the derivation of the main results presented above. Section \ref{Sec:Local} gives an analysis of the  correlations between the local shock sizes. Section \ref{Sec:Numerics} presents the results of our numerical analysis of these correlations for a toy model with a single degree of freedom,  i.e.\ $d=0$. Finally, a series of appendices contains technical derivations.

\section{Main results}\label{s:main-results}
Let us now state our main results for interfaces with a short-ranged elastic kernel (a more general case will be treated in the manuscript, with little changes to the  formulas). To this aim, we parameterize the position of
the interface    by the (real, one-component) displacement field $u(x)$, where
$x \in \mathbb{R}^d$ is the internal coordinate of the interface. For notational
convenience we     denote   $u(x) \equiv u_x$. The interface is submitted
to a quenched random potential $V(u_x,x)$, and to an external parabolic confining field $\frac {m^2}2 (u_x-w)^2$ centered at   $w$. 
In a given disorder realization $V$, upon variation of the external fiel $w$, the ground state (i.e.\ lowest-energy) configuration 
of the interface,
denoted $u_x(w)$, changes discontinuously  at a set of discrete locations $w_i$, according to   \be u_x(w_i^-) \to u_x(w_i^+) = u_x(w_i^-)+S^{(i)}_{x}\ .\ee The event $(w_i , S^{(i)}_{x})$ is the $i^{th}$ shock of the interface, $w_i$ is the location of the shock, $S^{(i)}_{x}$ is its local size at $x$ and $S^{(i)} = 
\int d^d x \, S^{(i)}_{x}$   its total size. The statistical properties associated to one shock were thoroughly analyzed using FRG in \cite{LeDoussalWiese2008c,LeDoussalWiese2011b}. Such properties are   encoded in the {\it shock density} $\rho_0$, defined as
\be 
\rho_0 := \overline{\sum_i \delta(w-w_i)} \ , 
\ee 
and in the {\it avalanche-size density} 
\be \label{defrhointro}
   \rho(S) := \overline{ \sum_{i} \delta(w-w_i) \delta(S-S^{(i)}) } \ . 
~~~~\ee 
The  {\it shock-size density} $\rho(S)$ is  linked to $\rho_0$ through $\rho_0 = \int dS \,\rho(S)$. 
Note that these quantities do not depend on $w$ due to the  statistical translational invariance (STS) of the disorder. Considering two points $w < w'$ and sizes $S_1 < S_2$, $ \int_{w}^{w'}d\tilde w \int_{S_1}^{S_2} d S \rho(S)$ is the mean number of shocks occurring between $w$ and $w'$ with   size $ S \in [S_1,S_2]$, while $(w'-w)\rho_0$ is the mean number of shocks (irrespective of their size). Note that throughout the rest of this section we will discuss our results in terms of densities but they can be translated into results for normalized probabilities as we discuss in Sec.~\ref{subsec:Probabilities}.

These observables alone do not  determine the statistical properties of the    sequence $\{(w_i , S^{(i)})\}_{i \in \mathbb{Z}}$ of shocks experienced by the interface in a given environment. In particular, they do not contain any information about the correlations between the shocks. For a given distance $W>0$, let us therefore introduce the {\it two-shock density at distance $W, $}
\be 
\rho_2(W):= \overline{ \sum_{i\neq j } \delta(w-w_i)   \delta(w+W-w_j) }  \ .
\ee  
This observable scales as {\it the square of a density}. Thus $\int_{w_1}^{w_1'} dw \int_{w_2}^{w_2'} dw' \rho_{2}(w'-w)$ counts {\it the mean number of pairs  of shocks} such that the first shock occurs between $w_1$ and $w_1'$, and the second one between  $w_2$ and $ w_2'$. Equivalently, $\tilde{\rho}_2(W) := \frac{\rho_2(W)}{\rho_0}$ is the density of shocks at a distance $W$ from a given shock. These observables   contain information about the correlations between shocks. Indeed an uncorrelated sequence of shocks implies $\rho_2(W) = \rho_0^2$ (and thus $\tilde{\rho}_2(W) = \rho_0$). A central question addressed in this work is whether the presence of a shock at a given point decreases ($\rho_2(W) < \rho_0^2$) or increases ($\rho_2(W) > \rho_0^2$) the density of shocks at a distance $W$.

To measure   the correlations between the size of the shocks (and not only their positions)
we   introduce the   {\it two-shock size density} at distance $W$, 
\bea
&&\!\!\!\!\!\!\! \rho_W(S_1,S_2):= \label{6}   \\
&& \overline{ \sum_{i\neq j } \delta(w-w_i) \delta(S_1-S^{(i)})  \delta(w+W-w_j) \delta(S_2-S^{(j)}) } \nn  \ .
\eea
It  is  linked to $\rho_2(W)$ via 
\be
\rho_2(W) = \int dS_1 \,dS_2 \,\rho_W(S_1,S_2)\ .\ee
Here $\int_{w_1}^{w_1'} dw \int_{w_2}^{w_2'} dw' \int_{S_1}^{S_1'} dS \int_{S_2}^{S_2'} dS' \rho_{w'-w}(S,S')$ counts the mean number of pairs of shocks such that the first  shock occurred between $w_1$ and $w_1'$, and the second between $w_2$ and $w_2'$, with sizes between $S_1$ and $S_1'$, resp. $S_2$ and $S_2'$. For this observable, an absence of correlations in the sequence of shocks   implies $\rho_W(S_1 , S_2) = \rho(S_1) \rho(S_2)$. To investigate the presence of correlations we thus study the {\it connected two-shock size density} $\rho_W^c(S_1,S_2)$, defined as
\be  \label{7}
\rho_W^c(S_1,S_2) := \rho_W(S_1,S_2) - \rho(S_1) \rho(S_2)\ .
 \ee 
At the level of mean-field theory, i.e.\ in the BFM model, it is known \cite{LeDoussalWiese2011b,ThieryLeDoussalWiese2015} that the shocks are independent and the process $w \to u_x (w)$ is a Levy jump process. As a consequence,  $\rho_W^c(S_1,S_2) =0$. On the other hand, for realistic interface models below their upper critical dimension,  the shocks are correlated, demanding to go beyond the BFM. This can   be seen from the second moment for which we show below the   {\it exact} relation 
\be\label{S1S2exactIntro}
 \frac{\langle S_1 S_2 \rangle_{\rho_W^c} }{[\langle
S \rangle_\rho]^2} =    - \frac{\Delta''(W)}{L^d m^4}\ .
\ee
On the left-hand-side,   $\langle
... \rangle_{\rho_W^c}$ denotes the average with respect to 
$\rho_W^c$ as defined in Eq.\ (\ref7).
 On the right-hand-side, $L$ is the lateral extension of the system, and $m^2$ the curvature of the confining potential, which sets the correlation length $ L_m := 1/m$ for avalanches in the lateral direction.  Finally,  $\Delta(W)$ is the  renormalized disorder-force correlator, the central object in the FRG treatment of disordered elastic systems: Denoting $u(w)$ the center-of-mass position of the interface, given well-position $w$,  the correlator $\Delta(W)$ is defined as the connected correlation function of the center-of-mass fluctuations of the interface position \cite{LeDoussal2006b}, 
\be
\Delta(W):= L^d m^4 \overline{ \left[ u(w) -w\right] \left[ u({w+W})-(w+W)\right] }^c\ .
\ee Up to a universal scaling factor and a single non-universal scale, the function $\Delta(W)$ only depends on the universality class of the problem. It was computed up to two-loop accuracy in Ref.~\cite{LeDoussalWieseChauve2003} and measured numerically in Ref.~\cite{MiddletonLeDoussalWiese2006}. For our purpose it is   important  that the function $\Delta(W)$ is uniformly of order $\epsilon$, and that its second derivative is non-zero. 
Thus the correlations (\ref{S1S2exactIntro}) increase when 
%
%
 going away from the upper critical dimension, where  mean-field theory, or equivalently  the BFM is relevant. Indeed, for the BFM    $\Delta''(W) = 0$,  and  the effective disorder force is distributed as a Brownian motion. Beyond mean-field theory, the   sequence of shocks  is correlated, thus  the effective disorder force at large scales has a different statistics than  Brownian motion. The sign of these correlations depends   on the sign of $\Delta''(W)$, which, in turn, depends on the universality class of the problem. As detailed in Sec.~\ref{SubSec:Delta},  {\it our results predict    qualitatively different   correlations depending on the universality class.}  The most important static universality classes of non-periodic, short-ranged disorder are  the random-bond (RB)   universality class, which at the microscopic level has short-ranged potential-potential correlations, and the random-field (RF) universality class, for which  the   force-force correlations, but not the potential-potential correlations, are short-ranged at the microscopic level. As is  summarized in Fig.~\ref{fig:Cartoon}, for RF-disorder $\Delta''(W)>0$, and thus avalanches are always anti-correlated. On the other hand, for  RB-disorder, avalanches are anti-correlated at short distances $W$, but  positively correlated at larger ones. 
\begin{figure}
\centerline{\includegraphics[width=4.5cm]{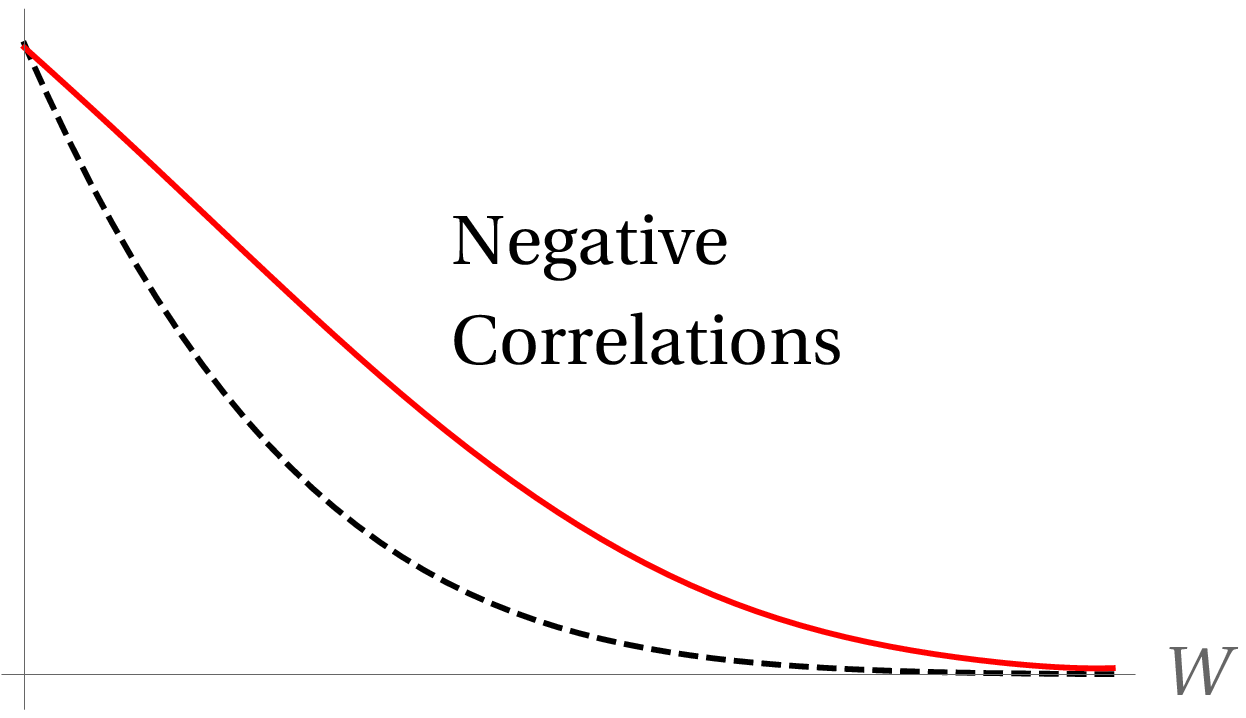} \includegraphics[width=4.5cm]{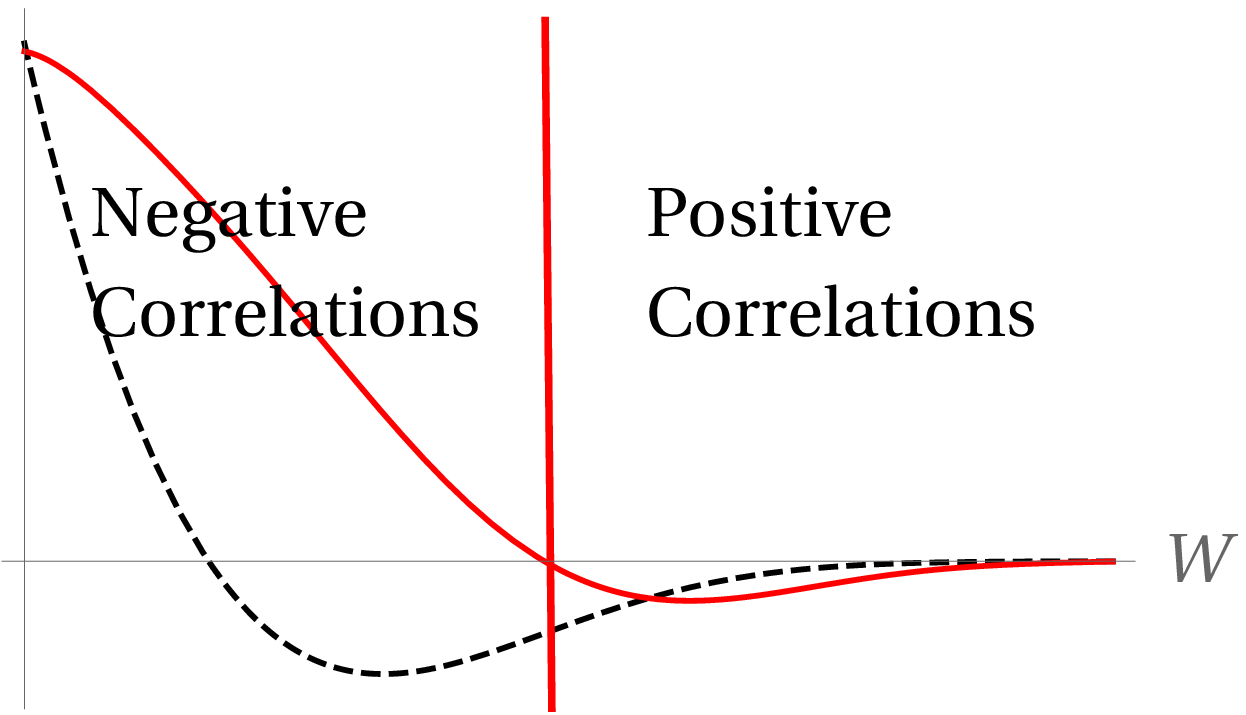}} 
\caption{Cartoons of the typical shape of the renormalized disorder correlator $\Delta(W)$ (black-dashed line) and of its second order derivative $\Delta''(W)$ (red line) for the Random-Field (left) and Random-Bond (right) universality classes (not to scale). Our results predict that the shock sizes are always negatively correlated in the Random-Field universality class, whereas the Random-Bond universality class exhibits a richer structure with negatively (resp. positively) correlated shock sizes at small (resp. large) distances.}
\label{fig:Cartoon}
\end{figure}

To obtain results for higher avalanche-size moments, we   use the FRG and the $\epsilon=(d_{\rm uc}-d)$ expansion   to show that, to lowest non-trivial order in the expansion, 
\be \label{MainResult}
 \rho^c_W(S_1,S_2)=   -  \frac{\Delta''(W)}{L^d m^4} \frac{S_1 S_2}{4S_m^2}  \rho(S_1)\rho(S_2) + O(\epsilon^2)\ .
\ee
Here  \be
S_m:= \frac{\langle S^2 \rangle_{\rho}}{2  \langle S \rangle_{\rho}}    \  , 
\ee
where $\langle
... \rangle_{\rho}$ denotes the average with respect to $\rho$ as defined in Eq.~(\ref{defrhointro}), is the characteristic size of avalanches, which acts as a large-scale cutoff for the avalanche-size density $\rho(S),$ and $\Delta''(W)$   introduced above is $O(\epsilon)$.
 Integrating Eq.~(\ref{MainResult}) times $S_1 S_2$ over $S_1$ and $S_2$, we recover Eq.~(\ref{S1S2exactIntro}). Contrary to the latter equation which is exact, relation (\ref{MainResult}) is correct only to order $\epsilon$.

As a consequence of Eq.~(\ref{MainResult}), and its generalizations to higher order, the correlations between avalanches
are universal. To make this more transparent, we rewrite Eq.\ (\ref{MainResult})  as
\be 
 \rho_W^c(S_1,S_2) = \frac{1}{(L m)^d} \frac{L^{2d}}{S_m^4} 
 {\cal F}_d\Big(\frac{W}{W_m}, \frac{S_1}{S_m} , \frac{S_2}{S_m}\Big) \label{eps} \ .
\ee  
The function ${\cal F}_d$ is universal and apart from its three arguments depends only on the spatial dimension.
To first order in $d=d_{\rm uc} -\epsilon$, and in the limit of large $L$ and small $m$, it is given by 
\be
 {\cal F}(w, s_1 , s_2) \simeq \frac{A_d \tilde  \Delta^{* \prime \prime}(w)}{16 \pi \sqrt{s_1 s_2} } e^{-(s_1+s_2)/4} 
 + O(\epsilon^2)\ . 
\ee 
 Here $A_d$ is an explicit constant, 
with $A_{d=4}=8 \pi^2$ for SR elasticity; the scale $W_m \sim m^{-\zeta}$, with $\zeta $ the roughness exponent contains a non-universal amplitude.  The range of validity of this result is discussed in  the main text. The presence of the  factor of $1/(Lm)^d$  highlights the fact that the correlations between shocks are local (indeed $N:=(Lm)^d$    counts the number of elastically independent regions of the interface). We will analyze  this local structure by studying the correlations between the local sizes of the shocks. 

To summarize,  let us  emphasize again our main message namely that for realistic models (beyond mean-field) the sequence of shocks   is always correlated.

\section{Model, shock observables and method} \label{Sec:Model}

\subsection{Model}
Consider the Hamiltonian for a $d-$dimensional elastic interface with position   $u(x) \equiv u_x \in \mathbb{R}$ ($ x \in \mathbb{R}^d$), elastic  kernel $g_{xx'}^{-1}$, subjected to a harmonic well centered at $w, $ and to a disorder potential $V(u,x)$:
\be \label{Hamiltonian1}
{\cal H} [u;w] = \frac{1}{2} \int_{xx'} g_{xx'}^{-1}(u_x - w)(u_{x'}-w) + \int_x V(u_x,x) \ .
\ee
Here $\int_x = \int d^dx$ and we assume everywhere that the system is confined in a box of length $L$ with e.g.\ periodic boundary conditions (the boundary conditions will not play a role in the following). We also assume the existence of a short-scale length cutoff  $a$. The elastic kernel is 
translationally invariant ($g_{xx'}^{-1} = g_{x-x'}^{-1}$) and defines a convex elastic-energy functional (i.e.\ $g_{xx'}^{-1}>0$ for $x \neq x'$). We denote $g_{q}^{-1} = 1/g_q$ its Fourier transform defined as $g_q^{-1} = \int_q e^{iq x} g_{x}^{-1}$, where $\int_q = \int \frac{d^d q}{(2\pi)^d}$. A possible choice is the standard short-ranged elasticity defined by \be \label{short-rangedKernel}
g_{xx'}^{-1} = \delta_{xx'} (-\nabla_{x'}^2 + m^2) \quad , \quad g_{q}^{-1} = q^2+m^2
\ .\ee
Here $\delta_{xx'}$ is the Dirac $\delta$ distribution, and the elastic coefficient has been set to one using an appropriate choice of units. Another   kernel   we   consider is  
\be  \label{long-rangedKernel}
g_q^{-1} = (q^2+\mu^2)^{\frac{\gamma}{2}}\ ,
\ee 
where $\gamma =2 $ corresponds to the previous case, and   $\gamma=1$ is       relevant for long-ranged elasticity, as  encountered in fracture and contact-line experiments. For a kernel of the form (\ref{long-rangedKernel}) we define the mass term as
\be \label{defmass}
m^2 := g^{-1}_{q=0} = \mu^{\gamma} .
\ee
It is the strength of the harmonic well. For short-ranged elasticity we have
\bea
{\cal H}_{\rm el} [u;w] &  :=& \frac{1}{2}\int_{xx'} g_{xx'}^{-1}(u_x - w)(u_x'-w)  \nn \\
&=&   \frac{1}{2}\int_{x} (\nabla_x u_x)^2 + m^2(u_x - w)^2 .
\eea
Thus $L_m:=m^{-1}$   defines a length scale beyond which different parts of the interface are elastically independent. It 
also provides a  large-scale cutoff in loop integrals encountered in the field theory. For more general kernels (\ref{long-rangedKernel}) this length scale is $L_{\mu}:= \mu^{-1}$, and we suppose $L_\mu \ll L$, ensuring that   boundary conditions do not play a role. The number of elastically independent parts of the interface is $N = (L/L_{\mu})^d$. The disordered potential $V(u,x)$ is assumed to be short-ranged in internal space $x$, and statistically translationally invariant, with a second cumulant
\be 
\overline{ V(u,x) V(u',x')}^c = \delta_{xx'} R_0(u-u') \ .
\ee  
The overline $\overline{(...)}$   denotes the average   over the disorder, and superscript $c$ stands for connected averages.  The detailed form of $R_0$ is, apart from   global features that determine the universality class of the problem (see Sec.~\ref{SubSec:Delta}), unimportant. We   also consider the force-force cumulant $\Delta_0(u) = - R_0''(u)$ such that $\overline{ \partial_u V(u,x)  \partial_{u'} V(u',x')}^c = \delta_{xx'} \Delta_0(u-u') $. Introducing a (finite) temperature $T$, disorder and thermal averages in this model can efficiently be computed using a replicated field theory. Introducing $n$ replicated fields $u_{ax}$, $a=1 , \dots , n $, the replicated action reads
\bea \label{Action1}
S[u] &=&  \frac{1}{2 T} \sum_{a} \int_{xx'}  g_{xx'}^{-1}(u_{ax} - w)(u_{ax'}-w) \nn \\
&&  - \frac{1}{2T^2}  \sum_{a,b} \int_x R_{0}(u_{ax} - u_{bx}) + \cdots 
\eea
where $\cdots$ indicates eventual higher cumulants of the disorder.

\subsection{The ground state and the scaling limit}
As discussed in the introduction, 
we are interested in the minimal energy configuration of the interface for a given parabolic well position $w$ and disorder realization $V$(i.e.\ the $T=0$ problem). It is defined as the configuration $u_x(w)$, which minimises the energy, 
\be 
u_x(w) := \underset{u_x}{{\rm argmin}} ~ {\cal H} [u;w] \ .
\ee 
We   denote
\be 
u(w) := \frac{1}{L^d}\int_x u_x(w),
\ee 
the center of mass of the ground-state of the interface. The statistical properties of $u_x(w)$ have been extensively studied in the literature. In particular it is known that the interface is self-affine with a (static) roughness exponent $\zeta$, defined by $\overline{[u_x(w) - u_{x'}(w)]^2} \sim |x-x'|^{2 \zeta}$. This scaling form generally holds in the scaling regime $L_c \ll |x-x'| \ll  L_{\mu}$ where $L_c$ is the Larkin length. The scaling limit is thus obtained for $L_{\mu} \to \infty$ or equivalently for $\mu \to 0$, also equivalent to $m \to 0$ (see (\ref{defmass})), a regime which is 
implicit throughout this work. In the FRG treatment of this problem, the ground state statistics  is studied using  the replicated field theory (\ref{Action1}). The mass term $m$ (or $\mu = m^{2/\gamma}$) can be conveniently used as a control parameter to study the flow of the effective action. As $m\to 0$ and through a proper rescaling, the effective action 
approaches 
a RG fixed point. This fixed point is perturbative in $\epsilon = d_{{\rm uc}}-d >0$ where $d_{{\rm uc}}$ is the upper critical dimension of the model (for kernels of the form (\ref{long-rangedKernel}) it is given by $d_{{\rm uc}}=2 \gamma$, thus $d_{{\rm uc}}=4$ for short-ranged elasticity and $d_{{\rm uc}}=2$ for long-ranged elasticity). The central object of the theory is the effective disorder correlator $R(u)$, a renormalized version of $R_0(u)$. It appears in the effective action of the theory $\Gamma[u],$ as $R_0(u)$ appears in the bare action $S[u]$ of Eq.~(\ref{Action1}) (see the action (\ref{Action3}) below). Remarkably, as shown in Ref.\ \cite{LeDoussal2008}, it is related to a physical observable, the renormalized disorder force-force   correlator $\Delta(u)$ defined as
\be  \label{DeltaDef}
\Delta(w-w'):= L^d m^4\overline{[u(w) -w][u(w') -w']}^c \ ,
\ee 
through the relation $\Delta''(u) = -R(u)$. This is the function that appears in the results (\ref{S1S2exactIntro}) and (\ref{MainResult}) of the introduction. The RG flow can be equivalently studied on $R$ or $\Delta$. For $m \to \infty$, the correlator $\Delta(w)$ is equal to the bare force-force correlator: $\Delta(w ) \to_{m\to \infty} \Delta_0(w)$. In the limit $m \to 0$ it admits a scaling form
\be  \label{DeltaRescaled}
 \Delta(w)  = A_d \mu^{\epsilon - 2 \zeta} \tilde \Delta (\mu^{\zeta} w)
\ee 
where $A_d$ is a dimensionless constant, and we recall $\mu = m^{2/\gamma}$. For kernels of the form (\ref{long-rangedKernel}), a convenient choice is to take $A_d$ as $A_d = \frac{1}{\epsilon \tilde{I}_2}$ with the dimensionless loop integral $\tilde{I}_2 := \int_q \frac{1}{(1+q^2)^{\gamma}}$. Note that the combination $\epsilon \tilde{I}_2$ stays finite as $\epsilon \to 0$. In general 
\be  \label{1loopIntegral}
A_d^{-1}= \epsilon \tilde{I}_2 = \frac{2}{(2 \sqrt{\pi})^d} \frac{\Gamma(\gamma+1-d/2)}{\Gamma(\gamma)}
\ ,\ee  and for example $\epsilon \tilde{I}_2 =_{\gamma=2; d= 4} 1/(8 \pi^2) $ and $ \epsilon \tilde{I}_2=_{\gamma=1; d = 2} 1/(2 \pi)$. As $m \to 0$, the rescaled disorder correlator $\tilde \Delta$ converges to  the fixed point of the FRG flow equation $\tilde{\Delta}^*(u)$, which  depends only on the universality class.

Let us now recall some important properties of these fixed-point functions.

\subsection{Properties of $\tilde{\Delta}^*(u)$ and static universality classes} \label{SubSec:Delta}

Depending on the properties of the bare disorder correlator $R_0(u)$, the FRG predicts that $\tilde \Delta(u)$ converges as $m \to 0$ to one of the fixed point of the FRG equation. A property of the (zero-temperature)
FRG equation is that, for non-periodic disorder, if $\tilde{\Delta}^*(u)$ is a fixed point, $\kappa^2 \tilde{\Delta}^*(u/\kappa)$ also is   a fixed point. Hence the fixed point towards which the system flows contains one non-universal scale whose value depends on microscopic properties of the disorder. The known fixed points can be regrouped into four main classes\footnote{There are other classes with different long-range correlations, but we will not
study them.}. Analytic properties of these fixed-point functions are known up to two-loop order, i.e.\ $O(\epsilon^2)$,  see Ref.\ \cite{LeDoussalWieseChauve2003} to which we refer the reader for quantitative results. 
An important property is that   all fixed points exhibit a cusp around $0$, $\Delta(u) \simeq \Delta(0) + \Delta'(0^+) |u| + O(u^2)$,  
related to the presence of avalanches \cite{LeDoussalWiese2008c,LeDoussalWiese2012a}. For our analys  the sign of $(\Delta^*)''(u)$ is crucial   as it determines the sign of the correlations. From the exact result (\ref{S1S2exactIntro}) (shown below) we see that for $(\tilde{\Delta}^*)''(W)>0$ shock sizes  at distance $W$  are anti-correlated, whereas for  $(\tilde{\Delta}^*)''(W)<0$) they are  positively correlated.

\paragraph*{Random-bond:} This class has a bare disorder potential $V(x,u)$ distributed with short-ranged correlations in the $u$ direction: The bare disorder correlator $R_0(u)$ decays quickly to $0$ as $u \to \infty$. The most important property for our analysis of the fixed-point function $\tilde{\Delta}^*_{{\rm RB}}(u)$ (its typical form is plotted on the right of Fig.~\ref{fig:Cartoon}) is that
$(\tilde{\Delta}^*_{{\rm RB}})''(u)>0$ at small $u$ and $(\tilde{\Delta}^*_{{\rm RB}})''(u)<0$ at large $u$.

\paragraph*{Random field:} This class has the bare disorder force $F(x,u)=-\partial_u V(x,u)$ distributed with short-ranged correlations. Then the bare force-force correlator $\Delta_0(u)$ is short-ranged and $R_0(u) \simeq_{u \gg 1} - \sigma |u|$ where $\sigma$ is called the amplitude of the random field. The most important property for our analysis of the fixed point function $\tilde{\Delta}^*_{{\rm RF}}(u)$ (its typical form is plotted on the left of Fig.~\ref{fig:Cartoon}) is that
$(\tilde{\Delta}^*_{{\rm RF}})''(u)>0$ for all $ x$.

\paragraph*{Random periodic:} This class corresponds to periodic disorder $V(u+1) = V(u)$. As a consequence,  $\tilde{\Delta}^*(u)$ is also periodic and  $(\tilde{\Delta}^*)''(u)=(\tilde{\Delta}^*)''(0)>0$ is constant. Though our analysis still applies to this universality class and our results are correct to $O(\epsilon)$, we will not discuss it here. As the
shock process is periodic in any dimension,   correlations  naturally arise from this periodicity (in particular in 
$d=0$ in the $m \to 0$ limit only one shock survives per interval).

\paragraph*{The Brownian-Force-Model universality class:} Finally, the Brownian-Force-Model defined as $\Delta_0(u) = - \sigma |u|$ is also a fixed point of the FRG flow equation and attracts all bare disorder such that  $\Delta_0(u) \simeq  - \sigma' |u|$ at large $u$. It   models avalanches at the mean-field level. (It resums tree diagrams). 
In this model   shocks are uncorrelated.

Hence, from the perspective of practical applications, the {\it qualitative} behavior of the correlations between shocks as a function of the distance   strongly depends on the universality class of the model (see Fig.~\ref{fig:Cartoon}).

\subsection{Shocks observables: Densities}

As recalled in the introduction, it is well known that in the limit of small $m$ the (rescaled) ground state $u_x(w)$ is piecewise constant as a function of $w$. In terms of the sequence of shocks $\{(w_i , S^{(i)}_{x})\}_{i\in \mathbb{Z}}$ one can write $u_x(w)$ and $u(w)$ as
\bea \label{ShockDecomp}
 u_x(w) &=& \sum_{i} \theta(w-w_i) S^{(i)}_{x}  \ , \nn \\
 u(w) &=& \frac{1}{L^d} \sum_{i} \theta(w-w_i) S^{(i)}  \ , 
\eea
where $\theta(x)$ is the Heaviside theta function. We recall   the definition of the one and two-shock size-density:
\begin{align}
& \rho(S) = \overline{ \sum_{i} \delta(w-w_i) \delta(S-S^{(i)}) }  \ ,  \\
& \rho_W(S_1,S_2) = \nn\\
& \overline{ \sum_{i\neq j } \delta(w-w_i) \delta(S_1-S^{(i)})  \delta(w+W-w_j) \delta(S_2-S^{(j))}) } \ .  \nn \\
\end{align}
 These distributions possess a large-scale cutoff which we denote $S_m$; the latter diverges for $m$   to $0$ as $S_m \sim m^{  - d - \zeta}$. Additionally, we suppose that they have a small-scale cutoff $S_0$. In the scaling regime,  $\rho(S)$ behaves as a power law with a characteristic exponent $\tau$: $\rho(S) \sim S^{-\tau}$ for $S_0 \ll S \ll S_m$. We us also define the connected density
\be 
\rho_W^c(S_1,S_2) = \rho_W(S_1,S_2) - \rho(S_1) \rho(S_2) \ .
\ee 
In the first part of this work our goal is to compute $\rho_W^c(S_1,S_2)$ up to first order in $\epsilon$ using the FRG.

\subsection{Shocks observables: Probabilities} \label{subsec:Probabilities}

One can normalize the above densities to define proper probability distributions as follows:
\bea 
  \rho_0 &:=& \int  \rho(S)\, dS\ , \\
 \rho_2(W) &:=& \int  \rho_W(S_1,S_2)\, dS_1 \,dS_2  \ , \\
 P(S) &:=& \frac{\rho(S)}{\rho_0}\ ,   \\
 P_W(S_1,S_2) &:=& \frac{\rho_W (S_1,S_2)}{\rho_2(W)} \label{defproba} \ .  
\eea
With this definition, $\rho_0 dw$ is the mean number of avalanches occurring in an interval $dw$ and $\int_{w_1}^{w_2}dw\int_{w_3}^{w_4}dw' \rho_2(w'-w)$ counts the number of pairs of shocks where the first   one occurs between $w_1$ and $w_2$ and the second between $w_3$ and $w_4$, irrespective of their sizes.  Given these definitions, $P(S)$ and $P_W(S)$ are normalized probability distribution functions (PDF). $ \int_{S}^{S'} d\tilde S\, P(\tilde S)$ is the probability, given that a shock has occurred, that its size is between $S$ and $S'$. $ \int_{S_1}^{S_1'} dS \int_{S_2}^{S_2'} dS' P_W(S,S')$ is the probability, given that two shocks occurred at a distance $W$, that their sizes are between $S_1$ and $S_1'$, and $S_2$ and $S_2'$. Note that a priori the marginal distribution$\int dS_1 P_W(S_1,S_2)$ is different from $P(S_2)$ since it contains the additional information that a shock occurred at a distance $W$. At the level of these PDFs, the absence of correlations would imply $P_W(S_1,S_2) = P(S_1,S_2)$ and, though in the remaining of the text we will favor the use of densities, our results can be translated to probabilities using Eq.~(\ref{defproba}). As discussed in Ref.~\cite{LeDoussalWiese2008c}, for an avalanche-size distribution $\rho(S)$ with exponent $\tau>1$ (which is relevant here), the value of $\rho_0$ is dominated by the small-scale cutoff  $S_0$ for avalanche sizes,  and diverges as $S_0 \to 0$,
\be 
\rho_0 = \int_{S_0}^{\infty} \rho(S) dS \sim_{S_0 \to 0} S_0^{1-\tau}  \ .
\ee 
Hence, $\rho_0$ is non-universal. In the same way $\rho_2(W)$ is non-universal, even though its relation with $\rho_0$ has some universal features   as we will show below. We denote by $\langle... \rangle_{\rho}$, $\langle... \rangle_{\rho_W}$, $\langle... \rangle_{\rho_W^c}$, $\langle... \rangle_{P}$ and $\langle... \rangle_{P_W}$ the averages with respect to $\rho$, $\rho_W$, $\rho_W^c$, $P$ and $P_W$.

\subsection{Relation between avalanche-size moments and renormalized force cumulants: First moment} \label{Subsec:FirstMom}
The $n^{\rm th}$ cumulant of the renormalized pinning force is defined as\bea \label{forceCumul}
 &&\!\!\!\!m^{2 n } \overline{ [u(w_1)-w_1] \dots [u(w_n)-w_n] }^c = \nn \\
 &&\qquad (-1)^n L^{-(n-1)d} \hat C^{(n)}(w_1,\dots,w_n)
\ .\eea
By definition $ \hat C^{(2)}(w_1,w_2) = \Delta(w_1-w_2)$ as   introduced above.    By parity invariance of the disorder $m^2\overline{[u(w)-w]} = 0,$ and thus $ \hat C^{(1)}(w)=0$.

\paragraph*{First cumulant:} One immediately gets by inserting Eq.~(\ref{ShockDecomp}) into $m^2\overline{[u(w)-w]} = 0$ the  exact relation 
\be 
\langle S \rangle_{\rho} = \rho_0 \langle S \rangle_{P} = L^d \ .
\ee 
\paragraph*{Second cumulant:} Differentiating with respect to $w_1$ and $w_2$ the definition $L^{-d} \Delta(w_1 - w_2) = m^4 \overline{[u(w_1)-w_1][u(w_2)-w_2]}$ with Eq.~(\ref{ShockDecomp}) inserted, one obtains the relation (33) of \cite{LeDoussalWiese2008c} (with a corrected misprint $1 \to -1$). It can be written in the form
\begin{align} \label{S1S2exact0}
- \frac{\Delta''(w_1-w_2)}{L^d m^4} = &\, L^{-2d} \langle S^2 \rangle_{\rho} \delta(w_1-w_2) \nn \\
& +  L^{-2d} \langle S_1 S_2 \rangle_{\rho_{w_2-w_1}}  - 1  \ . 
\end{align}
Hence, as pointed out in Ref.~\cite{LeDoussalWiese2008c}, the singular part of the second derivative of $\Delta''(w_1-w_2)$ around $w_2 = w_1$ gives an exact relation between the cusp in the renormalized disorder correlator
 \be  \label{defsigma}
 \sigma := -\Delta'(0^+) = R'''(0^+)\ ,
 \ee 
 and the  second avalanche-size  moment,
\be 
S_m := \frac{ \langle S^2 \rangle_{\rho} }{2  \langle S \rangle_{\rho}} =  \frac{ \langle S^2 \rangle_{P} }{2  \langle S \rangle_{P}} = \frac{ \sigma}{m^4} .
\ee 
The avalanche size $S_m$ plays the role of a large-scale cutoff for $\rho(S)$. On the other hand, the regular part of Eq.~(\ref{S1S2exact0}) gives the exact relation
\be  \label{S1S2exact}
L^{-2d} \langle S_1 S_2 \rangle_{\rho_W} = 1 - \frac{\Delta''(W)}{L^d m^4} .
\ee 
For uncorrelated shocks we would have obtained $L^{-2d} \langle S_1 S_2 \rangle_{\rho_W} = 1$. The correlations thus come from the non-zero value of $\Delta''(W) \neq 0$, a property which is generally expected from the FRG. It is a simple signature of the fact that the effective disordered force felt by the interface at large scale is not Brownian. Note that in terms of the moments of the connected density, the exact relation (\ref{S1S2exact}) reads
\be  \label{S1S2exact2}
L^{-2d} \langle S_1 S_2 \rangle_{\rho_W^c} = - \frac{\Delta''(W)}{L^d m^4} .
\ee 
Let  us also write   the exact relation (\ref{S1S2exact}) in terms of the probabilities defined in Sec.~\ref{subsec:Probabilities}:
\bea
\frac{\rho_2(W)}{\rho_0^2} \frac{\langle S_1 S_2 \rangle_{P_W} }{(\langle S \rangle_P)^2} = 1 - \frac{\Delta''(W)}{L^d m^4}   . 
\eea

\subsection{Generating functions}

We now introduce the generating functions which encode   all the moments of the density $\rho_W(S_1,S_2)$. Let us first recall the generating functions used in the one-shock case:
\bea \label{Z1shock1}
&& Z(\lambda ) = L^{-d} \langle e^{\lambda S} -1\rangle_{\rho} \ , \nn \\
&& \hat Z(\lambda) =  L^{-d} \langle e^{\lambda S} -\lambda S - 1\rangle_{\rho} = Z(\lambda)- \lambda  .
\eea
They are   related to observables associated with the position   as
\bea \label{Z1shock2}
&& Z(\lambda ) = L^{-d} \lim_{\delta \to 0^+} \partial_{\delta} \overline{ e^{ L^d[u(w+\delta) - u(w) ]}}  \ , \nn \\
&& \hat Z(\lambda) = L^{-d} \lim_{\delta \to 0^+} \partial_{\delta}  \overline{ e^{ L^d[\hat{u}(w+\delta) - \hat{u}(w) ]}} \ ,
\eea
where $\hat{u}(w) := u(w) - w$ is the translated position field. Note that due to STS they
are independent of $w$.
These relations were proven in Ref.~\cite{LeDoussalWiese2011b}. For   two shocks we introduce
\be  \label{Z2shocks1}
Z_{W}(\lambda_1,\lambda_2 ) := L^{-2d}\langle (e^{\lambda_1 S_1} -1)(e^{\lambda_2 S_2} -1) \rangle_{\rho_W} \ .
\ee 
We show in Appendix \ref{app:Gene} that it can be computed as
\bea \label{Z3}
\lefteqn{ Z _W(\lambda_1, \lambda_2 )} \nn\\
&=& \hat{Z} _W (\lambda_1 , \lambda_2 ) + \lambda_2 \hat{Z}(\lambda_1) + \lambda_1 \hat{Z}(\lambda_2) + \lambda_1 \lambda_2 \nn \\
&=&  \hat{Z} _W (\lambda_1 , \lambda_2 ) +\lambda_2 Z(\lambda_1) + \lambda_1 Z(\lambda_2) - \lambda_1 \lambda_2  \ .~~~~~~~~
\eea
We used the definition
\bea \label{Z2shocks2}
&& \!\!\!\!\!\!\!\!\!\!\! \hat Z_{w_2-w_1}(\lambda_1,\lambda_2 ) := L^{-2d} \times  \\
&& \!\!\!\!\!\!\!\!\!\!\! \lim_{\delta_1, \delta_2 \to 0^+} \partial_{\delta_1 , \delta_2} \overline{ e^{ L^d  \lambda_1[\hat u(w_1+\delta_1) - \hat u(w_1)  ]} e^{ L^d  \lambda_2[\hat u(w_2+\delta_2) - \hat u(w_2)  ]}} \nn
\eea
In the following we   compute $\hat Z_W(\lambda_1 , \lambda_2)$ using the FRG through   formula (\ref{Z2shocks2}). Let us also define the connected generating functions
\bea 
Z_W^c(\lambda_1 , \lambda_2)  & :=& L^{-2d} \langle (e^{\lambda_1 S_1} -1)(e^{\lambda_2 S_2} -1) \rangle_{\rho_W^c} \nn \\
& =& Z_W(\lambda_1 , \lambda_2)- Z(\lambda_1)Z(\lambda_2) \nn \\
\hat Z_W^c(\lambda_1 , \lambda_2) & :=& \hat Z_W(\lambda_1 , \lambda_2)- \hat Z(\lambda_1) \hat Z(\lambda_2)  \label{defhatZc} 
\eea
These functions are actually equal: $Z_W^c(\lambda_1 , \lambda_2)=\hat Z_W^c(\lambda_1 , \lambda_2)$ as is easily seen using~(\ref{Z3}).

\subsection{Relation between avalanche-size moments and renormalized force cumulants: Kolmogorov cumulants and chain rule}
Using Eq.\ (\ref{Z2shocks2}) and the fact that $\overline{\hat u(w)}=0$, the generating function $\hat Z_W(\lambda_1 , \lambda_2)$ can be written
as\bea \label{expand} 
\lefteqn{\hat Z_W(\lambda_1 , \lambda_2) =  \sum_{n,m=1}^{\infty} \frac{\lambda_1^n \lambda_2^m }{n! m!} \lim_{\delta_1, \delta_2 \to 0^+} } \\
 && \frac{L^{(n+m-2 )d }}{\delta_1 \delta_2} \overline{[\hat u(\delta_1) - \hat u(0)  ]^n [\hat u( W +\delta_2) - \hat u(W)  ]^m} \ .\nn
\eea
In the limit of $\delta_i \to 0$    we   encounter for each $(n,m)$ two types of terms:
\bea \label{connVsdisc}
\lefteqn{ \overline{[\hat u(\delta_1) - \hat u(0)  ]^n [\hat u( W +\delta_2) - \hat u(W)  ]^m} = } \nn \\
&& \overline{[\hat u(\delta_1) - \hat u(0)  ]^n}^c \times \overline{[\hat u(W+\delta_2) - \hat u(W)  ]^m}^c   \\
&& + \overline{[\hat u(\delta_1) - \hat u(0)  ]^n [\hat u( W +\delta_2) - \hat u(W)  ]^m}^c + O(\delta_i^3) \ .\nn 
\eea
The term in the second line of Eq.~(\ref{connVsdisc}) produces the disconnected part of the avalanche moment $\langle S_1^{n} \rangle \langle S_2^{m} \rangle$ and thus the disconnected part of the generating function $\hat Z_W(\lambda_1 , \lambda_2)$, that is $\hat Z(\lambda_1) \hat Z(\lambda_2)$. The last term on the other hand     contributes to $\langle S_1^{n} S_2^{m} \rangle_{\rho_W^c}$ and to the connected part of the generating function, $\hat Z_W^c(\lambda_1 , \lambda_2)=Z_W^c(\lambda_1 , \lambda_2)$ which is the true unknown. Introducing the Kolmogorov cumulants
\bea \label{KolmoCum}
\lefteqn{ K_W^{(n,m)}(\delta_1 , \delta_2) := } \\
&& L^{(n+m-2 )d} \overline{[\hat u(\delta_1) - \hat u(0)  ]^n [\hat u( W +\delta_2) - \hat u(0)] ^m}^c  \ ,\nn
\eea
we can write
\be
Z_W^c(\lambda_1, \lambda_2) =  \sum_{n,m=1}^{\infty} \frac{\lambda_1^n \lambda_2^m}{n! m!} \lim_{\delta_1, \delta_2 \to 0^+}  \frac{1}{\delta_1 \delta_2} K_W^{(n,m)}(\delta_1 , \delta_2)  \ , 
\ee
or, equivalently,
\be  \label{chainRule}
   \langle S_1^{n} S_2^m \rangle_{\rho_W^c} = \lim_{\delta_1, \delta_2 \to 0^+} \frac{1}{\delta_1 \delta_2} K_W^{(n,m)}(\delta_1 , \delta_2) \ .
\ee 
The Kolmogorov cumulants (\ref{KolmoCum}) can be generally extracted from the renormalized force cumulants (\ref{forceCumul}), as we now explain. Let us   introduce\footnote{Note that those differ from ${\cal C}$ introduced in \cite{LeDoussalWiese2008c} by an additional factor of $L^{-d}$.}
\bea \label{mom1}
&&\!\!\!\! C^{(n,m)}(w_1 , \dots ,w_n , w_{n+1},  \dots ,w_{n+m}) =  \\
&& \quad L^{(n+m-2 )d} \overline{\hat{u}(w_1) \dots \hat{u}(w_n) \hat{u}(w_{n+1}) \dots \hat{u}(w_{n+m})   }^c  \ . \nn 
\eea
They are trivially linked to the renormalized force cumulants (\ref{forceCumul}): $C^{(n,m)}(w_1 , \dots ,w_n , w_{n+1},  \dots ,w_{n+m}) = \frac{1}{L^d} (-1/m^2)^{n+m} \hat C^{(n+m)}(w_1 , \dots w_{n+m})$. 
Explicit expressions for the lowest cumulants with $n+m \leq 4$ are displayed in Ref.\ \cite{LeDoussalWiese2008c},
see e.g.\ Eq.\ (61) there.
In the notation for $C^{(n,m)}$, though the expression is symmetric in $w_{i}$, we have highlighted the facts that in the end the $n$ first $w_i$ will be taken around $w=0$, whereas the last $m$  will be around $W$. Indeed,  to obtain   $K^{(n,m)}_{W} (\delta_1 , \delta_2)$ from the moments $C^{(n,m)}$, we must successively evaluate $C^{(n,m)}$ with $w_i \to \delta_1$ minus $C^{(n,m)}$ with $w_i \to 0$ for each $i = 1 , \dots ,n$, then set $w_i \to W+\delta_2$ minus $C^{(n,m)}$ with $w_i \to W$ for each $i = n+1 , \dots ,n+m$. Ambiguities associated with the possible presence of terms such as $\Delta'(0^{\pm})$, are lifted by taking the limit of coinciding points with a given specific ordering of the $w_i$. Consistency requires that the end result does not depend on the   chosen ordering, a property   linked to the assumption that all singularities of
the field $\hat u(w)$ can be modeled by a finite density of dilute shocks (which
guarantees e.g.\ the continuity of   $\hat C$).
 This iterative procedure was called the ${\cal K}$ operation in \cite{LeDoussalWiese2008c}.

\subsection{Strategy of the calculation and validity of the results} \label{subsec:Strategy}

In order to compute $\hat Z_W(\lambda_1 , \lambda_2)$, we must be able to perform disorder averages of moments of the position field at various positions $w_i$ for $i=1 , \dots , r$. For example $r=4$ is sufficient in the formulation (\ref{Z2shocks2})  and used in Appendix \ref{app:Algebraic}. In the main part of this work we report a calculation of $\hat Z_W(\lambda_1 , \lambda_2)$ from the study of the moments (\ref{mom1}) and we   thus need to keep $r$ arbitrary. We therefore consider the theory for $r$ position fields $u_{x}^i$ coupled to different parabolic wells centered at positions $w_i$ in the same disordered environment. The Hamiltonian of the problem is
\be 
{\cal H}[\{u\}, \{w \}] = \sum_{i=1}^{r} {\cal H}_{\rm el}[u^i, w^i] + \sum_{i=1}^r \int_x V(u_x^i , x) \ .
\ee 
This leads to a replicated action of the form
\bea \label{Action2}
S[u] &=&  \frac{1}{2 T} \sum_{a,i} \int_{xx'}  g_{xx'}^{-1}(u_{ax}^i - w_i)(u_{ax'}^i-w_i) \nn \\
&&  - \frac{1}{2T^2}  \sum_{a,i; b ,j} \int_x R_{0}(u_{ax}^i - u_{bx}^j) + \cdots 
\eea
The effective action of the theory is \cite{LeDoussalWiese2008c,LeDoussalWiese2011b,LeDoussal2008}
\bea \label{Action3}
\Gamma[u] &=&  \frac{1}{2 T} \sum_{a,i} \int_{xx'}  g_{xx'}^{-1}(u_{ax}^i - w_i)(u_{ax'}^i-w_i) \nn \\
&&  - \frac{1}{2T^2}  \sum_{a,i; b ,j} \int_x R(u_{ax}^i - u_{bx}^j) + O(\epsilon^2)
\ .~~~~~\eea
Here $R(u) = O(\epsilon)$ is the renormalized disorder correlator already introduced in the previous section, while the neglected terms are higher-order terms in $\epsilon$ that can be expressed as loop integrals with higher powers of $R$. The calculation of observables using the effective action (\ref{Action3}) has been called the {\em\ improved tree approximation} \cite{LeDoussalWiese2008c,LeDoussalWiese2011b}. Here we did not specify the number of replicas $a=1 , \dots , n_r$. As is usual in replica calculations, the $n_r \to 0 $ limit will be implicit in the following. Since (\ref{Action3}) is the effective action, observables will be computed using a saddle-point calculation, or equivalently in a diagrammatic language, by resuming all tree diagrams generated by the action (\ref{Action3}). This calculation   allows to get the lowest order in $\epsilon$ for any observable. Let us recall   the known results at the improved tree level for $\rho(S)$ and $Z(\lambda)$ as obtained in Refs.\ \cite{LeDoussalWiese2008c,LeDoussalWiese2011b}:
\bea \label{1shockResult}
 \rho(S) &=& \frac{L^d}{2 \sqrt{\pi} S^{\frac{3}{2}} (S_m)^{\frac{1}{2}}} e^{-\frac{S}{4S_m}} \ ,  \\
 Z(\lambda)&=& \lambda + S_m Z(\lambda)^2 = \frac{1}{2 S_m} (1 - \sqrt{1 - 4 \lambda S_m}) \ . ~~~~~~~
\eea

\subsection{Connected versus non-connected averages and the $\epsilon$-expansion}

 Before going further, let us now mention a subtle point. As will become clear in the following, the improved tree calculation leads to a result of order $O(\epsilon)$ for $\rho_W^c$, in contrast to $\rho(S)$ for which it leads to a result of order $O(1)$ \footnote{To be   rigorous, this is only true of the dimensionless density $\tilde \rho(\tilde S) = S_m^2 \rho(S_m \tilde S)$ since $S_m = O(\epsilon)$, we neglect this subtlety in the following.}. Hence if one computes $\rho_W(S_1,S_2) =\rho(S_1) \rho(S_2)  + \rho_W^c(S_1,S_2)  $ to $O(\epsilon)$ one must pay attention to the fact that $\rho_W^c(S_1,S_2)$ can be computed using the improved-tree theory, but $\rho(S)$ has then to be computed to one-loop accuracy. In the same way, the connected generating function
 \be 
 Z_W^c(\lambda_1 , \lambda_2) = Z_W(\lambda_1 , \lambda_2) - Z(\lambda_1) Z(\lambda_2)
 \ee 
can be computed exactly up to order $O(\epsilon)$ using the improved tree theory, but to compute $Z_W(\lambda_1 , \lambda_2) $ up to order $\epsilon$ one must   add one-loop corrections to $Z(\lambda)$. The same remark holds for the moments $\langle S_1^{n_1} S_2^{n_2} \rangle_{\rho_W^c} = \langle S_1^{n_1} S_2^{n_2} \rangle_{\rho_W}-\langle S_1^{n_1} \rangle_{\rho}\langle S_2^{n_2} \rangle_{\rho}$.

\section{Correlations between total shock  sizes}\label{Sec:Global}

\subsection{Reminder of the diagrammatic rules and extraction of shock moments} 
Let us now explain how the moments 
\bea \label{mom1bis}
\lefteqn{C^{(n,m)}(w_1 , \dots ,w_n , w_{n+1},  \dots ,w_{n+m})  } \\
&=& L^{(n+m-2 )d} \overline{\hat{u}(w_1) \dots \hat{u}(w_n) \hat{u}(w_{n+1}) \dots \hat{u}(w_{n+m})   } \nn \\
&=&  L^{-2d} \int_{y_1 \dots y_{n+m}} \overline{\hat{u}_{y_1}(w_1) \dots  \hat{u}_{y_{n+m}}(w_{n+m})   } \nn
\eea
are obtained using the diagrammatic rules developed in Ref.~\cite{LeDoussalWiese2008c} which can also be read off from the action (\ref{Action3}). In the calculation of the correlator (\ref{mom1}), the terms of the form $L^d \hat{u}(w_i) = \int_{y_i} \hat u_{y_i}(w_i)$ are diagrammatically represented as external legs at the top of the diagrams. Fields at different position $w_i$ and $w_j$ can be contracted through an interaction vertex $\int_z \frac{1}{T^2} R(\hat u_z(w_i) - \hat u_z(w_j) + w_i - w_j)$, represented as a dashed-line (each contraction bringing an additional derivative to $R$ with the appropriate sign). The propagators are represented as plain lines. When forming tree diagrams, one produces $n+m -1$ interaction vertices $\frac{1}{T^2} R,$ and $2(n+m-1)$ propagators, which each carries a factor of $T$. For trees, all factors of $T$ cancel,  and the diagrams survive in the $0$ temperature limit. The factors of $T$  can thus   be omitted in the  diagrammatic rules. As for the integrals over the   positions  of the external legs $y_i$, $i=1 , \dots , n+m$ and the disorder vertices $z_k$, $k=1 , \dots , n+m-1$, since the interaction is local in space and $\int_{x}g_x = \frac{1}{m^2}$, all   $2(n+m-1)$ propagators can be taken as static propagators and thus this integration  produces an additional factor of $L^d$. 
This procedure results in expressions for the $C^{(n,m)}(w_1 , \dots ,w_n , w_{n+1},  \dots ,w_{n+m})$ as sums of products of terms involving derivatives $\Delta^{(p)}(w_i-w_j)$ \footnote{Each $R$ vertex must be contracted at least twice or there would be one free-replica sum left in the replicated theory, leading to $0$ in the limit of a vanishing number of replicas.}. 
In calculating the Kolmogorov cumulants $K^{(n,m)}(\delta_1,\delta_2)$ to order $O(\delta_1 \delta_2)$ 
one must use the even but non-analytic form of $\Delta(u)$ around the origin,
\be 
\Delta(u) = \Delta(0) + \Delta'(0^+) |u| + \frac{\Delta''(0)}{2} u^2 + O(u^3)\ .
\ee 
We checked  that  if one takes all limits of coinciding points  with a fixed order  of the $w_i$ in the calculation, one obtains a non-ambiguous result, independent
of the   ordering.

\subsection{Lowest moments}

\paragraph*{First moment:} We fist consider the computation of $\langle S_1 S_2 \rangle_{\rho_W^c}$. To this aim we   compute $C^{(1,1)}(w_1 ,w_2)$, which is given by a single diagram:
\vspace{0.1cm}
\begin{eqnarray} \label{SecondMomDiagram}
C^{(1,1)}(w_1 ,w_2) &=&  \hspace{1cm} \begin{gathered}
\begin{fmffile}{secondMom}
\begin{fmfgraph*}(30,20)
\fmfleft{i1,i2}
\fmfright{o1,o2}
\fmf{plain, label=$ \frac{1}{m^2} $ }{i2,i1}
\fmf{dbl_dashes, label=$ \Delta(w_1-w_2)$ }{i1,o1}
\fmf{plain, label=$ \frac{1}{m^2} $}{o1,o2}
\fmflabel{$w_1  \approx 0 $}{i2}
\fmflabel{$w_2  \approx W$}{o2}
\end{fmfgraph*}
\end{fmffile}
\end{gathered} \nn \\
&&    \nn \\
&& \nn \\
&=&    \frac{1}{L^d m^4} \Delta(w_1-w_2) \ .
\end{eqnarray}
We have introduced   a new diagrammatic notation: A double-dashed line represents an  interaction vertex between position fields at a finite distance $ \approx W$;  we reserve the single dashed line for interaction vertices between nearby position fields. Hence,
\bea \label{K11}
K_W^{(1,1)}(\delta_1 , \delta_2) & =& \frac{1}{L^d m^4} \Big[ \Delta(-\delta_1 + W + \delta_2) - \Delta(  W + \delta_2)   \nn \\
&&  ~~~~~~~~ ~- \Delta(-\delta_1+ W) + \Delta(W) \Big]   \nn \\
& =& - \frac{\Delta''(W)}{L^d m^4} \delta_1 \delta_2 + O(\delta_i^2)  \ .
\eea
Using (\ref{chainRule}) we conclude that
\be  \label{S1S2epsilon}
L^{-2d} \langle S_1 S_2 \rangle_{\rho_W^c} = - \frac{\Delta''(W)}{L^d m^4}  \ .
\ee 
This is the exact result (\ref{S1S2exact}), here retrieved diagrammatically within the improved tree approximation. 
A priori there could be higher-order corrections $O(\epsilon^2)$ on the r.h.s.\ of (\ref{S1S2epsilon}), coming from
loop diagrams. However, the definition (\ref{DeltaDef}) of $\Delta(u)$ as a physical observable effectively resums an infinite number of loop diagrams. The same diagrams then arise on both sides of Eq.\ (\ref{S1S2epsilon}), and the result (\ref{S1S2exact}) is exact.

\medskip

\paragraph*{Second moment:} Let us now consider the computation of $\langle S_1^2 S_2\rangle_{\rho_W^c}$. We first need to compute $C^{(2,1)}(w_1,w_2,w_3)$. Diagramamtically it is given by
\begin{widetext}
\bea \label{thirdMomDiagram}
&& C^{(2,1)}(w_1,w_2,w_3)   \nn\\ \nn\\
&& = 2\, {\rm Sym}_{w_1 \leftrightarrow w_2} \left( \hspace{0.25cm}  \begin{gathered}
\begin{fmffile}{thirdMom}
\begin{fmfgraph*}(50,30)
\fmfleft{i1,i2}
\fmfforce{0w,1h}{i2}
\fmfforce{0w,0h}{i1}
\fmfforce{0.5w,1h}{o2}
\fmfforce{0.5w,0h}{o1}
\fmfforce{1w,0.5h}{o3}
\fmfforce{0.5w,0.5h}{v1}
\fmfforce{1w,1h}{o4}
\fmf{plain }{i2,i1}
\fmf{dashes}{i1,o1}
\fmf{plain}{o1,v1,o2}
\fmf{dbl_dashes}{v1,o3}
\fmf{plain}{o3,o4}
\fmflabel{$w_2  \approx 0 $}{i2}
\fmflabel{$w_1  \approx 0$}{o2}
\fmflabel{$w_3 \approx W$}{o4}
\end{fmfgraph*}
\end{fmffile}
\end{gathered}  
 \hspace{1.5cm} + \hspace{1.5cm} \begin{gathered}
\begin{fmffile}{thirdMom2}
\begin{fmfgraph*}(50,30)
\fmfleft{i1,i2}
\fmfforce{0w,1h}{i2}
\fmfforce{0w,0h}{i1}
\fmfforce{0.5w,1h}{o2}
\fmfforce{0.5w,0h}{o1}
\fmfforce{1w,0.5h}{o3}
\fmfforce{0.5w,0.5h}{v1}
\fmfforce{1w,1h}{o4}
\fmf{plain }{i2,i1}
\fmf{dbl_dashes}{i1,o1}
\fmf{plain}{o1,v1,o2}
\fmf{dashes}{v1,o3}
\fmf{plain}{o3,o4}
\fmflabel{$w_3 \approx W$}{i2}
\fmflabel{$w_1  \approx 0$}{o2}
\fmflabel{$w_2  \approx 0$}{o4}
\end{fmfgraph*}
\end{fmffile}
\end{gathered} \hspace{1.5cm}  +  \hspace{1.5cm} \begin{gathered}
\begin{fmffile}{thirdMom3}
\begin{fmfgraph*}(50,30)
\fmfleft{i1,i2}
\fmfforce{0w,1h}{i2}
\fmfforce{0w,0h}{i1}
\fmfforce{0.5w,1h}{o2}
\fmfforce{0.5w,0h}{o1}
\fmfforce{1w,0.5h}{o3}
\fmfforce{0.5w,0.5h}{v1}
\fmfforce{1w,1h}{o4}
\fmf{plain }{i2,i1}
\fmf{dbl_dashes}{i1,o1}
\fmf{plain}{o1,v1,o2}
\fmf{dbl_dashes}{v1,o3}
\fmf{plain}{o3,o4}
\fmflabel{$w_2 \approx 0$}{i2}
\fmflabel{$w_3  \approx W $}{o2}
\fmflabel{$w_1 \approx 0$}{o4} 
\end{fmfgraph*}
\end{fmffile} 
\end{gathered} \hspace{1cm} \right) \nn  \\
&& \nn \\
&& = \frac{2 }{L^d m^8} {\rm Sym}_{w_1 \leftrightarrow w_2}\Big[ \Delta(w_1-w_2)\Delta'(w_1-w_3) + \Delta(w_1-w_3)\Delta'(w_1-w_2)   + \Delta(w_3-w_2)\Delta'(w_3-w_1) \Big]   
\eea
\end{widetext}
In doing the ${\cal K}$ operation to go from $C^{(2,1)}$ to $K^{(2,1)}_W$, these diagrams are not equivalent. At   order $\delta_1 \delta_2$ that we are interested in, the first term leads to $4 \frac{\Delta'(0^+)}{m^4}\frac{\Delta''(W)}{L^d m^4} \delta_1\delta_2$, the second to $2 \frac{\Delta'(0^+)}{m^4}\frac{\Delta''(W)}{L^d m^4} \delta_1\delta_2$, whereas the third one is of order $O(\delta_1^2 \delta_2)$ and does not contribute. Using Eq.~(\ref{chainRule}) we conclude that
\be  \label{resS12S2}
L^{-2d} \langle S_1^2 S_2 \rangle_{\rho_W^c} =     6 \frac{\Delta'(0^+)}{m^4} \frac{ \Delta''(W)}{L^d m^4} + O(\epsilon^2) \ .
\ee 
\paragraph*{General rules for diagrams:} The last example is rather instructive for the three general rules: 

(i) the only diagrams that contribute to the Kolmogorov cumulant $K_W^{(n,m)}(\delta_1 , \delta_2)$ at order $\delta_1 \delta_2$ contain a single double-dashed   vertex (that is a single disorder interaction vertex connecting the two disjoint sets of points at $w \approx 0$ and $w \approx W$); 

(ii) this vertex becomes a $\Delta''(W)$ at order $\delta_1 \delta_2$; 

(iii) the other interaction vertices are between (almost) coinciding points, and  produce a factor of $\Delta'(0^+)$   at order $\delta_1 \delta_2$. These rules are discussed in Appendix~\ref{app:DiagramRule}. As a result,  diagrams contributing to the two-shock moments consist of diagrams reminiscent of the one-shock case (i.e.\ they contain only $\Delta'(0^+)$ vertices) linked together by an interaction vertex $- \frac{\Delta''(W)}{L^d m^4}$.

\subsection{Generating function for all moments}

Let us now use the above rules and give a diagrammatic computation of $Z_W^c(\lambda_1,\lambda_2)=\hat Z_W^c(\lambda_1 , \lambda_2) $ defined in Eq.~(\ref{defhatZc}).
To this aim, let us first introduce a diagrammatic notation for $Z(\lambda)$ defined in Eq.~(\ref{Z1shock1}):
\begin{equation} \label{Zdiagram}
Z(\lambda) = \begin{gathered}
\begin{fmffile}{Z1}
\begin{fmfgraph*}(50,30)
\fmfforce{0.25w,1h}{i2}
\fmfforce{0.25w,0.3h}{i1}
\fmfforce{0.75w,1h}{i3}
\fmfforce{0.75w,0.3h}{i4}
\fmfforce{0.650w,0.9h}{e1}
\fmfforce{0.35w,0.9h}{e2}
\fmfforce{0.5w,0.3h}{o1}
\fmf{plain }{i2,i1}
\fmf{plain }{i3,i4}
\fmf{dots}{e1,e2}
\fmfblob{0.5w}{o1}
\end{fmfgraph*}
\end{fmffile}
\end{gathered} \ .
\end{equation}
We have emphasized using dots that there is an arbitrary number of external legs at the top of the diagrams summed in Eq.~(\ref{Zdiagram}). Using the expansion (\ref{expand})
and following the rules explained in the previous section, the diagrams entering in $\hat Z_W^c(\lambda_1 , \lambda_2) $ are made of two trees linked by a single doubled dashed line.  It is the sum of all   tree diagrams  for  avalanches at $w=0$, times all   tree diagrams for avalanches at $w=W$, linked together by a single $- \frac{\Delta''(W)}{L^d m^4}$ inserted between any   pair of 
points belonging to each tree. This can be represented as
\vspace{0.1cm}
\begin{eqnarray} \label{ZWdiagram}
  Z_W^c(\lambda_1,\lambda_2)&=&\hat Z_W^c(\lambda_1 , \lambda_2) \nn\\\nn\\ \nn\\ 
& =& \begin{gathered}
\begin{fmffile}{ZW}
\begin{fmfgraph*}(50,30)
\fmfforce{0.25w,1h}{i12}
\fmfforce{0.25w,0.3h}{i11}
\fmfforce{0.75w,1h}{i13}
\fmfforce{0.75w,0.3h}{i14}
\fmfforce{0.650w,0.9h}{e11}
\fmfforce{0.35w,0.9h}{e12}
\fmfforce{0.5w,0.3h}{o11}
\fmf{plain }{i12,i11}
\fmf{plain }{i13,i14}
\fmf{dots}{e11,e12}
\fmfblob{0.5w}{o11}

\fmfforce{1.25w,1h}{i22}
\fmfforce{1.25w,0.3h}{i21}
\fmfforce{1.75w,1h}{i23}
\fmfforce{1.75w,0.3h}{i24}
\fmfforce{1.650w,0.9h}{e21}
\fmfforce{1.35w,0.9h}{e22}
\fmfforce{1.5w,0.3h}{o21}
\fmf{plain }{i22,i21}
\fmf{plain }{i23,i24}
\fmf{dots}{e21,e22}
\fmfblob{0.5w}{o21}

\fmfforce{0w,-0.9h}{i32}
\fmfforce{0w,-1.6h}{i31}
\fmfforce{0.5w,-0.9h}{i33}
\fmfforce{0.5w,-1.6h}{i34}
\fmfforce{0.4w,-1h}{e31}
\fmfforce{0.1w,-1h}{e32}
\fmfforce{0.25w,-1.6h}{o31}
\fmf{plain }{i32,i31}
\fmf{plain }{i33,i34}
\fmf{dots}{e31,e32}
\fmfblob{0.5w}{o31}

\fmfforce{0w,-0.9h}{i32}
\fmfforce{0w,-1.6h}{i31}
\fmfforce{0.5w,-0.9h}{i33}
\fmfforce{0.5w,-1.6h}{i34}
\fmfforce{0.4w,-1h}{e31}
\fmfforce{0.1w,-1h}{e32}
\fmfforce{0.25w,-1.6h}{o31}
\fmf{plain }{i32,i31}
\fmf{plain }{i33,i34}
\fmf{dots}{e31,e32}
\fmfblob{0.5w}{o31}

\fmfforce{1.5w,-0.9h}{i42}
\fmfforce{1.5w,-1.6h}{i41}
\fmfforce{2w,-0.9h}{i43}
\fmfforce{2w,-1.6h}{i44}
\fmfforce{1.9w,-1h}{e41}
\fmfforce{1.6w,-1h}{e42}
\fmfforce{1.75w,-1.6h}{o41}
\fmf{plain }{i42,i41}
\fmf{plain }{i43,i44}
\fmf{dots}{e41,e42}
\fmfblob{0.5w}{o41}

\fmfforce{0.5w, -0.2}{x1}
\fmfforce{1.5w, -0.2}{x2}
\fmf{dbl_dashes}{i42,i33}
\fmf{plain}{x1,i33}
\fmf{plain}{x2,i42}
\fmfv{decor.shape=circle,decor.filled=full,
decor.size=0.05w}{i33}
\fmfv{decor.shape=circle,decor.filled=full,
decor.size=0.05w}{i42}

\fmfforce{0.5w , 1h}{lab1}
\fmfforce{1.15w , 1.1h}{lab2}
\fmfforce{0.27w , -0.39h}{lab3}
\fmfforce{1.57w , -0.39h}{lab4}
\fmflabel{$w  \approx 0$}{lab1}
\fmflabel{$w  \approx W$}{lab2}
\fmflabel{$w  \approx 0$}{lab3}
\fmflabel{$w  \approx W$}{lab4}

\end{fmfgraph*}
\end{fmffile}
\end{gathered}\hspace{2.5cm} \ . ~~~~~~~
\end{eqnarray}
\vspace{2cm}

\noindent
The diagrams above the point of insertion of   $\Delta''(W)$   on the left are given by $Z(\lambda_1)$. The terms below are all the diagrams in $Z(\lambda_1)$ with an arbitrary external leg selected, that is $\frac{d Z(\lambda_1)}{d \lambda_1}$. A similar contribution arises on the right-hand side. Hence we arrive at the result
\bea \label{Zcres}
Z_W^c(\lambda_1 , \lambda_2)  &=&  - \frac{\Delta''(W)}{L^d m^4} Z(\lambda_1) \frac{d Z(\lambda_1)}{d \lambda_1}  Z(\lambda_2) \frac{d Z(\lambda_2)}{d \lambda_2}  \nn \\
&& + O(\epsilon^2) 
\eea
In terms of $Z_W(\lambda_1 , \lambda_2)$ this result reads
\bea \label{Zres}
Z_W(\lambda_1 , \lambda_2)& =& Z(\lambda_1) Z(\lambda_2)   \\
&& - \frac{\Delta''(W)}{L^d m^4} Z(\lambda_1) \frac{d Z(\lambda_1)}{d \lambda_1}  Z(\lambda_2) \frac{d Z(\lambda_2)}{d \lambda_2} \ . \nn
\eea
It is correct to $O(\epsilon)$ if one takes into account the $O(\epsilon)$ corrections to $Z(\lambda)$. Expanding the result (\ref{Zcres}) one obtains the   moments $\langle S_1^nS_2^m\rangle_{\rho_W^c}$:
\bea \label{momentRes}
 \lefteqn{ \langle S_1^n S_2^m\rangle_{\rho_W^c }  =  -\frac{\Delta''(W)}{L^{3d} m^4} n! m!  }  \\
&&  \times\sum_{p = 0}^{n-1} \sum_{q = 0}^{m-1}\frac{ \langle S^{n-p} \rangle_{\rho} \langle S^{p+1} \rangle_{\rho}\langle S^{m-q} \rangle_{\rho} \langle S^{q+1} \rangle_{\rho} }{(n-p)!p!(m-q)!q!}   + O(\epsilon^2) \ . \nn
\eea 
The diagrammatic interpretation of this result is straightforward: to construct an arbitrary diagram contributing to $\langle S_1^n S_2^m\rangle_{\rho_W^c }$, one must first choose $p \leq n-1$ external legs on the left that will be below the point of insertion of   $- \frac{\Delta''(W)}{L^d m^4}$ (there must be at least one leg above this point of insertion). In the ${\cal K}$ operation, all those points lead to a term that   contributes to  $\langle S^p \rangle_{\rho}$. The combinatorial term accounts for the $C^{n}_p$ possible choices. Note that this result was derived using the heuristic diagrammatic rules developed in the preceding section. We observe that:

(i)  It correctly reproduces the results for the small-order moments (\ref{S1S2epsilon}) and (\ref{resS12S2}). We   checked that it leads to $\langle S_1^3 S_2 \rangle_{\rho_W^c}= - 60 \frac{\Delta''(W)}{L^d m^4} S_m^2$ and $\langle S_1^2 S_2^2 \rangle_{\rho_W^c}= - 27 \frac{\Delta''(W)}{L^d m^4} S_m^2$, which can also be derived from the expression for $\hat C^{(4)}(w_1,w_2,w_3,w_4)$ given e.g.\ in formula (61) of Ref.~\cite{LeDoussalWiese2008c}.

(ii) We give in Appendix \ref{app:C-D} an alternative  derivation of Eq.~(\ref{Zres}) that uses the Carraro-Duchon formalism  \cite{CarraroDuchon1998,LeDoussalWiese2011b}.

(iii) We give in Appendix \ref{app:Algebraic} a derivation using a saddle-point calculation within the effective action (\ref{Action3}). This also yields the local structure of correlations studied in Section~\ref{Sec:Local}.

\subsection{Results for the densities}

To infer $\rho_W$ from Eq.~(\ref{Zres}), we first note the identity $Z(\lambda) \frac{d Z(\lambda)}{d \lambda} = \frac{1}{2S_m} \frac{d}{d\lambda}\left( Z(\lambda) - \lambda \right)$, derived from the self-consistent equation (\ref{1shockResult}) for $Z(\lambda)$. Differentiating $L^{-d} \int dS (e^{\lambda S} -1 ) \rho(S) = Z(\lambda)$ with respect to $\lambda$ and using $\langle S \rangle_{\rho} =L^d$ yields
\be
L^{-d} \int dS (e^{\lambda S} -1 ) S \rho(S) =\frac{d}{d\lambda}\left[ Z(\lambda) - \lambda \right] \ . 
\ee
Finally, using Eqs.\ (\ref{Z2shocks1}) and (\ref{Zres}),
we   obtain
\be  \label{resRho}
\rho_W(S_1,S_2) = \rho(S_1)\rho(S_2) \left( 1 -  \frac{\Delta''(W)}{L^d m^4} \frac{S_1S_2}{4S_m^2}\right) \ .
\ee 
This is our main result for the two-schock density, already announced in Eq.~(\ref{MainResult}) of the introduction. It can be used to extract a variety of physical observables.

\paragraph*{Mean number of pairs of shocks:}
Integrating over $S_1$ and $S_2$, we obtain two equivalent formulas for $\rho_2(W)$:
\bea \label{resRho2}
\rho_2(W) & =& \rho_0^2 - \frac{\Delta''(W)}{L^d m^4} \frac{L^{2d}}{4S_m^2} \\
& = &\rho_0^2\left[1- \frac{\Delta''(W)}{L^d m^4} \left(\frac{\langle S \rangle_P}{2S_m} \right)^{\!2} \right] \nn  \ .
\eea
 Hence, although both $\rho_0$ and $\rho_2(W)$ are non-universal and dominated by the non-universal small avalanche size cutoff $S_0$ discussed in Sec.~\ref{subsec:Probabilities},   the connected density $\rho_2(W)-  \rho_0^2$ does not depend on $S_0$ and is universal.

\paragraph*{Normalized probability distribution:}
The above results allow us to express the probability distribution $P_W(S_1,S_2) = \frac{\rho_W(S_1,S_2)}{\rho_2(W)} $
 to $O(\epsilon)$ accuracy as
\bea
\lefteqn{ P_W(S_1,S_2) =} \nn \\
&&  P(S_1) P(S_2) \left[ 1 -  \frac{\Delta''(W)}{4 S_m^2 L^d m^4} \Big(S_1S_2 - \langle S \rangle_P^2\Big)  \right]\ .  ~~~~~
\eea

\paragraph*{Conditional probability distribution:}
Another PDF of interest is the conditional probability to have a shock  with amplitude $S_2$, given that there was a shock of amplitude $S_1$ at a distance $W$ before.  To $O(\epsilon)$ accuracy 
\bea
P_W(S_2|S_1) &=& \frac{P_W(S_1,S_2)}{ \int dS_2 P_W(S_1,S_2) } \\
& =& P(S_2) \left[1 - \frac{\Delta''(W) S_1}{4 S_m^2 L^d m^4} \Big(S_2 -  \langle S \rangle_P \Big) \right]  \ .\nn
\eea
Its mean value, normalized by $\left< S\right>_P$, is
\be 
\frac{\langle S_2 | S_1 \rangle}{\langle S \rangle_P} =  1 - \frac{\Delta''(W) S_1}{4 S_m^2 L^d m^4} \Big(2 S_m - \langle S \rangle_P\Big) \ .
\ee 

\paragraph*{Second shock marginal:}
The probability for the size $S_2$ of a
second shock at $W$, given that there was a shock at $0$, is
\bea
 P_W(S_2) &=& \int dS_1 P_W(S_1,S_2) \nn \\
& =&  P(S_2) \left[1 - \frac{\Delta''(W) \langle S \rangle_P }{4 S_m^2 L^d m^4}  
\Big( S_2 - \langle S \rangle_P\Big) \right]  \ . \nn\\
\eea 
The normalized mean value of the second shock is
\be \label{S2} 
\frac{\langle S_2 \rangle_W}{\langle S \rangle_P} =  1 - \frac{\Delta''(W) \langle S \rangle_P}{4 S_m^2 L^d m^4}  \Big(2 S_m - \langle S \rangle_P\Big) \ .
\ee 

\subsection{Analysis of the results} \label{subsec:Analysis}

\paragraph*{Sign of the correlations:} As discussed in Sec.~\ref{SubSec:Delta}, the sign of the correlations (positively or negatively correlated shock sizes) solely depends on the sign of $\Delta''(W)$, which depends on the distance $W$ and on the universality class of the problem. The above results thus unveil a rich phenomenology for the correlations as pictured in Fig.~\ref{fig:Cartoon}.

\paragraph*{Range of validity:} The result (\ref{Zcres}) was obtained in the framework of the $\epsilon$ expansion. The results for the connected part of the correlations are by definition the first non-zero terms in this expansion, since they were obtained within the improved tree approximation, and they appear at $O(\epsilon)$. As a perturbative result, it is by definition  controlled for $\epsilon \to 0$. For finite $\epsilon$, the predictions should be accurate as long as the corrections to the mean-field behavior are small. This is worth emphasizing, since the moments $\langle S_1^nS_2^m\rangle_{\rho_W}$ predicted by the formula (\ref{momentRes}) become negative for large $(n,m)$, signaling a breakdown of the improved tree approximation. This is also the case of the two-shock density computed at the improved tree level in Eq.~(\ref{resRho}) which becomes negative at large $S_i$. There the approximation is not controlled anymore since $O(\epsilon)$ corrections are   larger than the mean-field result. Let us  see  when this occurs: Using the simple estimate $\Delta''(W)\approx |\Delta'(0^+)|/W_{\mu}$, where $W_\mu$ is the length of order $\mu^{-\zeta}$ on which $\Delta(W)$ decays, see below, and $|\Delta'(0^+)|=m^4 S_m$, the bound $\rho(S_1,S_2)>0$ is violated if
\be
1 \lesssim \frac{S_m} {W_\mu \mu^{-d}}\times \frac1{(\mu L)^d}  \times \frac{S_1 S_2}{4S_m^2}  \  .
\ee
The first factor is a dimensionless number of order $O(\epsilon)$ near $d=d_{{\rm uc}}$. The second vanishes in the thermodynamic limit of $L\to \infty$. Thus the bound can only  be violated if $S_1 S_2/S_m^2$ compensates this factor. This can  only  be achieved if at least one of the avalanches is either   system-spanning, or   far out in the tail of the distribution, i.e. the bound is only violated for very unlikely events.

 Note however that the exact result (\ref{S1S2exact}) is protected from being negative since
\be
L^{-2d}\langle S_1 S_2 \rangle_{\rho_W} = 1 - \frac{\Delta''(W)}{L^d m^4} = \overline{\partial_w u(w) \partial_w u(W+w)}\ , 
\ee
and $\partial_w u(w)$ is always positive since $u(w)$ is monotonically increasing as a function of $w$. The latter  can be shown rigorously using a stability argument: Writing that $u_x(w)$ is a stable minimum of the Hamiltonian (\ref{Hamiltonian1}) implies for all $x$  two equations, namely  $\frac{\delta {\cal H}[u,w]}{\delta u(x)}=0$, and 
$\frac{\delta^2 {\cal H}[u,w]}{\delta u(x) \delta u(y)}\ge 0$. Specifying the second equation to $x=y$, we obtain 
\bea \label{80}
&& m^2[u_x(w) - w] + \partial_u V(u_x(w) , x) = 0 \ ,  \\
&& m^2 + \partial_u^2 V(u_x(w) , x) \geq 0 . \label{81}
\eea
Taking a derivative of Eq.~(\ref{80}) w.r.t.\ $w$, solving for $\partial_w u_x(w)$, and using Eq.~(\ref{81}) implies  
\be
\partial_w u_x(w) =\frac1{1 +m^{-2} \partial_u^2 V(u_x(w) , x)} \geq 0\ .
\ee


\paragraph*{Comparison with experiments and numerics:} Though our predictions rely on the analysis of the model (\ref{Hamiltonian1}), they were obtained using FRG and thus we expect Eqs.\ (\ref{Zcres}) and (\ref{momentRes}) to be valid for all models in the same universality class. All our results, namely
Eq. (\ref{momentRes}) and Eqs. (\ref{resRho})-(\ref{S2}), contain the combination 
$\frac{\Delta''(W)}{L^d m^4}$. On one hand it can be used to give a result to order $O(\epsilon)$ in 
the form of a universal function (see below). On the other hand all quantities entering the r.h.s of these
equations can be measured directly in an experiment or in a numerical simulation. Indeed we recall that
\be S_m:=\frac{ \langle S^2 \rangle_P}{ 2 \langle S \rangle_P} \equiv \frac{ \langle S^2 \rangle_{\rho}}{ 2 \langle S \rangle_{\rho}}
\ee
 and the combination 
\be  \label{85}
\frac{\Delta''(W)}{L^dm^4} = \partial_W^2 \overline{[u(w) -w][u(w+W) -w -W]}^c 
\ee 
can both be measured and do not require to know the mass   $m$ which might be hard to identify. The computation of this second derivative then gives a   precise characterization of the amplitude of the correlations through the exact formula (\ref{S1S2exact2}). The accuracy of the $\epsilon$ expansion and   universality can then be tested against 
the formulas given in the previous section.

\paragraph*{Universal function:} Using   rescaled quantities we can 
rewrite our main result as (see Eq.~(\ref{DeltaRescaled}) and Sec.~\ref{SubSec:Delta})
\be 
 \rho_W^c(S_1,S_2) = \frac{1}{(L \mu)^d} \frac{L^{2d}}{S_m^4} 
 {\cal F}_d\Big(\frac{W}{W_\mu}, \frac{S_1}{S_m} , \frac{S_2}{S_m}\Big) \label{eps2} 
\ee  
where the function ${\cal F}_d$ is universal and depends only on the space dimension.
To first order in $d=d_{\rm uc} -\epsilon$, it is given by 
\be
 {\cal F}(w, s_1 , s_2) \simeq \frac{A_d  \Delta^{* \prime \prime}(w)}{16 \pi \sqrt{s_1 s_2} } e^{-(s_1+s_2)/4} 
 + O(\epsilon^2) 
\ee 
in the limit of large $L$ and small $\mu$ and $A_d$ was given in Eq.~(\ref{1loopIntegral}). Here $ \Delta^{* \prime \prime}(w)$ is the universal fixed point of the FRG equation,  normalized to 
$\Delta^*(0)=\epsilon$. Indeed, for small $m$ the rescaled renormalized disorder correlator of the system $\tilde{\Delta}(w)$, appearing in Eq.~(\ref{DeltaRescaled}), is close to one of the fixed points of the FRG equation: $\tilde{\Delta}(w) \simeq \tilde{\Delta}^{*}(w)$. For non-periodic disorder, the latter can be expressed using one constant $\kappa$ as $\tilde \Delta^*(w)= \kappa^2 \Delta^*(w/\kappa)$ (see Sec.~\ref{SubSec:Delta}). The parameter $\kappa$ is thus the
single non-universal constant in our formula. The scales
in Eq. (\ref{eps2}) are then given by 
\be  \label{scalingWm}
W_{\mu} \simeq \kappa \mu^{-\zeta}  \  ,\qquad  S_m \simeq A_d \kappa  \Delta^{* \prime}(0^+) \mu^{-(d + \zeta)} 
\ee  
for small $\mu$. We remind that $m = \mu^{\gamma/2}$. We have defined all quantities such that their expressions are the most simple ones, independent of $\gamma$. With the above normalization, to order $\epsilon$,
$ \Delta^{* \prime}(0^+) = \sqrt{\epsilon(\epsilon-2 \zeta)}$ and
$ \Delta^{* \prime \prime}(0) = \frac{2 \epsilon}{9}$.

\paragraph*{Locality:} Note that in the result (\ref{eps2}) the amplitude 
of the correlation is inversely proportional to $N = (L\mu)^d$, the number of elastically independent degrees of freedom  of the interface. This is a signature of the local nature of the correlations. For two shocks a distance  $W$ apart, there is a probability of order $1/N$ that they occur in the same region of space. To go further 
into this locality property and to remove this bias 
we investigate in the next section the correlations between the local shock  sizes.

\section{Local structure of correlations} \label{Sec:Local}

In this section we analyze the correlations between the local shock  sizes. We start by deriving a general formula for the correlations between the local shock sizes measured on an arbitrary subset of   the internal space of the interface. To this aim we define
\be 
S_1^{\phi_1} = \int_x S_{1x} \phi_{1x} \  , \qquad S_2^{\phi_2} = \int_x S_{2x} \phi_{2x}\ ,
\ee 
where $\phi_1$ and $\phi_2$ are two arbitrary test functions. Two extreme cases are $\phi_{1x} =1$: in this case $S_1^{\phi_1} = S_1$, and the observable is the total size studied in the precedent section. The other extreme is    $\phi_{1x} = \delta^d(x-x_1)$, for which $S_{1}^{\phi_1} = S_{1x_1}$ is the local size at $x=x_1$.

\subsection{Reminder: one-shock case}
Here we briefly recall the essential definitions and results given in Refs.\ \cite{LeDoussalWiese2008c,LeDoussalWiese2011b} on the density and generating function associated to the local one-shock size statistics. For a general test function $\phi$ we introduce
\bea
 \rho^{\phi}(S^{\phi}) &:=& \overline{ \sum_i \delta ( S^{(i),\phi} - S^{\phi})  \delta(w_i-w) }  \ , \nn \\
 Z^{\phi}(\lambda) &:=& \frac{1}{\int_x \phi_x}\langle e^{\lambda S^{\phi}} -1 \rangle_{\rho^{\phi}}  \ , \nn \\
 \hat Z^{\phi}(\lambda) &:=& Z^{\phi}(\lambda) - \lambda \ ,
\eea
where $\langle ... \rangle_{\rho^{\phi}}$ denotes the average with respect to $\rho^{\phi}$. Note that   $\hat Z^{\phi}$ has no linear term,   since the first moment of $\rho^{\phi}$ is  due to  STS 
\be 
\langle S^{\phi} \rangle_{\rho^{\phi}} = \int_{x} \phi_x \ .
\ee 
The generating function $\hat Z^{\phi}(\lambda)$ is obtained from the replica field theory using the exact relation 
\be 
\hat Z^{\phi}(\lambda) = \frac{1}{\int_x \phi_x} \partial_{\delta} \overline{e^{\int_x \phi_x [u_x(w+\delta) -u_x(w) - \delta ] }}|_{\delta=0^+} \ .
\ee 
It was shown in Refs.~\cite{LeDoussalWiese2008c,LeDoussalWiese2011b} that $Z^{\phi}(\lambda)$ can be written as 
\be
Z^{\phi}(\lambda)= \frac{\int_x Z_x^{\phi} (\lambda)}{\int_x \phi_x}  \ , 
\ee
where, at the improved-tree-theory level, $Z^{\phi}_x(\lambda)$ satisfies the following self-consistent equation
\be  \label{local:selfcons}
Z^{\phi}_x(\lambda) = \lambda \phi_x + \sigma \int_{yy'} g_{x-y} g_{x-y'} Z^{\phi}_y(\lambda) Z^{\phi}_{y'}(\lambda) \ .
\ee 
The quantity  $\sigma = -\Delta'(0^+)$ was defined in Eq.~(\ref{defsigma}).

\subsection{Two-shock case: Notation and diagrammatic result}

\paragraph*{Densities and generating functions:} Consider
\begin{widetext}
\be 
   \rho_{W}^{\phi^1 \phi^2}(S_1^{\phi^1} , S_2^{\phi^2} ) :=   \overline{\sum_{i\neq j} \delta(w-w_i) \delta(S_1^{\phi^1}-S^{(i),\phi^1})  \delta(w+W-w_j) \delta(S_2^{\phi^2}-S^{(j),\phi^2}) }  \ . \nn
\ee 
The generating functions are 
\begin{align} \label{local:hatZ}
& Z_W^{\phi^1\phi^2} := \frac{1}{\int_x \phi^1_x\int_x \phi^2_x}\left< \left(e^{\lambda_1 S_1^{\phi^1}} -1\right)\left(e^{\lambda_2 S_2^{\phi^2}} -1\right) \right>_{\!\rho^{\phi^1\phi^2}_W}   \\
& \hat Z^{\phi^1\phi^2}_{w_2-w_1}(\lambda_1,\lambda_2 ) := \frac{1}{\int_x \phi^1_x\int_x \phi^2_x} \lim_{\delta_1, \delta_2 \to 0^+}  \partial_{\delta_1 , \delta_2} \overline{ e^{\int_x \phi^1_x \lambda_1[\hat u_x(w_1+\delta_1) - \hat u_x(w_1)  ]} e^{ \int_x \phi^2_x  \lambda_2[\hat u_x(w_2+\delta_2) - \hat u_x(w_2)  ]}}  \ ,  
\end{align}
\end{widetext} 
where $\langle ... \rangle_{\rho^{\phi^1\phi^2}_W}$ denotes the average with respect to $\rho_{W}^{\phi^1 \phi^2}$. The following relation holds
\bea \label{local:hatZandZ}
Z_W^{\phi^1\phi^2}(\lambda_1, \lambda_2 )& =&  \hat{Z}_W^{\phi^1\phi^2} (\lambda_1 , \lambda_2 ) + Z^{\phi^1}(\lambda_1)  \lambda_2 \nn \\
&&   + \lambda_1 Z^{\phi^2}(\lambda_2)  - \lambda_1 \lambda_2  \ .
\eea
(These relations are a consequence of  Appendix \ref{app:Gene}). The connected equivalents of the previous definitions are  constructed as in the previous section for the correlations between the total sizes; for example 
\begin{equation}
\rho^{c; \phi^1 \phi^2}_W (S_1^{\phi^1} , S_2^{\phi^2} ) = \rho^{ \phi^1 \phi^2}_W (S_1^{\phi^1} , S_2^{\phi^2} ) - \rho^{\phi^1 } (S_1^{\phi^1} )\rho^{\phi^2 }( S_2^{\phi^2} ),
\end{equation}
and   we   note $\langle ... \rangle_{\rho^{c;\phi^1\phi^2}_W}$ the average w.r.t.\  $\rho_{W}^{c;\phi^1 \phi^2}$.

\paragraph*{Simplified notation for averages:} In order that these somewhat complicated notations 
  do not obscure our results, 
we   introduce   simplified notations for averages. We first note that 
\be 
\rho_{W}^{\phi^1 \phi^2}(S_1^{\phi^1} , S_2^{\phi^2} ) = \rho_2(W) {\cal P}(S_1^{\phi^1} , S_2^{\phi^2} ),  
\ee 
where $\rho_2(W)$ is as before the density of a pair of shocks and   ${\cal P}(S_1^{\phi^1} , S_2^{\phi^2} )$ denotes the probability, given that two shocks occured at a distance $W$, that their local sizes measured with respect to $\phi^1$ and $\phi^2$ are $S_1^{\phi^1}$ and $S_2^{\phi^2}$. We have 
dropped the dependence of   ${\cal P}$ on $\phi^1$ and $\phi^2$ to alleviate our notations.
We also note arbitrary moments as
\bea
 \langle  \langle (S_1^{\phi1})^n (S_2^{\phi^2})^m \rangle \rangle_{\rho_W} &:=&  \langle (S_1^{\phi1})^n (S_2^{\phi^2})^m \rangle_{\rho_{W}^{\phi^1 \phi^2}}   \\
 \langle  \langle (S_1^{\phi1})^n (S_2^{\phi^2})^m \rangle \rangle_{\rho_W^c} &:=&  \langle (S_1^{\phi1})^n (S_2^{\phi^2})^m \rangle_{\rho_{W}^{c;\phi^1 \phi^2}} . ~~~~~~~
\eea
We indicate the dependence on the choice of $\phi^1$ and $\phi^2$ only inside the average, and not in the measure. A moment of the form $\langle  \langle (S_1^{\phi1})^n (S_2^{\phi^2})^m \rangle \rangle_{\rho_W}$ is thus equal to the product of $\rho_2(W)$ and of the mean value of $(S_1^{\phi1})^n (S_2^{\phi^2})^m$ for shocks at a distance $W$, given that two such shocks occurred.

\paragraph*{Diagrammatic result:} In Appendix \ref{app:Algebraic} we compute these generating functions by a direct evaluation of Eq.~(\ref{local:hatZ}) using a saddle-point calculation on the effective action (\ref{Action3}). Alternatively, from a diagrammatic point of view, the result can   be adapted from the reasoning that led to $ Z_{W}(\lambda_1,\lambda_2 )$ by {\it keeping track of the space dependence in the different vertices, propagators and sources in the diagram} (\ref{ZWdiagram}). Following Eq.~(\ref{Zdiagram}), we represent 
$Z_x^{\phi}(\lambda)$ as
\vspace{0.2cm}
\begin{equation} \label{ZLocdiagram}
Z_x^{\phi}(\lambda) =  \begin{gathered}
\begin{fmffile}{Z1loc}
\begin{fmfgraph*}(50,30)
\fmfforce{0.25w,1h}{i2}
\fmfforce{0.25w,0.3h}{i1}
\fmfforce{0.75w,1h}{i3}
\fmfforce{0.75w,0.3h}{i4}
\fmfforce{0.650w,0.9h}{e1}
\fmfforce{0.35w,0.9h}{e2}
\fmfforce{0.5w,0.3h}{o1}
\fmf{plain }{i2,i1}
\fmf{plain }{i3,i4}
\fmf{dots}{e1,e2}
\fmfblob{0.5w}{o1}
\fmfforce{0.5w , 1h}{lab1}
\fmfforce{0.5w , -0.12h}{lab2}
\fmfv{decor.shape=circle,decor.filled=full,
decor.size=0.07w}{lab2}
\fmflabel{$\phi$}{lab1}
\fmflabel{$x$}{lab2}
\end{fmfgraph*}
\end{fmffile}
\end{gathered} \ .
\end{equation}

\vspace{0.3cm}
\noindent
The same diagram without the marked point $x$  is also used to represent $\int_x Z_x^{\phi}(\lambda)$, itself equal to $\int_x \phi_x  \times Z^{\phi}(\lambda)$. Then, as before, $\hat Z^{\phi^1 \phi^2}_{W}(\lambda_1,\lambda_2 )$ is the sum of a connected and a disconnected part: 
\be\label{localDisc}
\hat Z^{\phi^1 \phi^2}_{W}(\lambda_1,\lambda_2 ) = \hat Z^{\phi^1}(\lambda_1)  \hat Z^{\phi^2}(\lambda_2)  + \hat Z^{c,\phi^1 \phi^2}_{W}(\lambda_1,\lambda_2 ) \ .
\ee 
The connected part $\hat Z^{c;\phi^1 \phi^2}_{W}(\lambda_1,\lambda_2 )$ is
\vspace{0.5cm}
\begin{eqnarray} \label{ZWlocaldiagram}
\lefteqn{  \hat Z^{c;\phi^1 \phi^2}_{W}(\lambda_1,\lambda_2 )} \nn\\
&&  =  \frac{1}{\int_x \phi^1_x  \int_x \phi^2_x }  \times  \rule[-25mm]{0mm}{25mm}{\begin{gathered}
\begin{fmffile}{ZWlocal}
\begin{fmfgraph*}(50,30)
\fmfforce{0.25w,1h}{i12}
\fmfforce{0.25w,0.3h}{i11}
\fmfforce{0.75w,1h}{i13}
\fmfforce{0.75w,0.3h}{i14}
\fmfforce{0.650w,0.9h}{e11}
\fmfforce{0.35w,0.9h}{e12}
\fmfforce{0.5w,0.3h}{o11}
\fmf{plain }{i12,i11}
\fmf{plain }{i13,i14}
\fmf{dots}{e11,e12}
\fmfblob{0.5w}{o11}

\fmfforce{1.25w,1h}{i22}
\fmfforce{1.25w,0.3h}{i21}
\fmfforce{1.75w,1h}{i23}
\fmfforce{1.75w,0.3h}{i24}
\fmfforce{1.650w,0.9h}{e21}
\fmfforce{1.35w,0.9h}{e22}
\fmfforce{1.5w,0.3h}{o21}
\fmf{plain }{i22,i21}
\fmf{plain }{i23,i24}
\fmf{dots}{e21,e22}
\fmfblob{0.5w}{o21}

\fmfforce{0w,-0.9h}{i32}
\fmfforce{0w,-1.6h}{i31}
\fmfforce{0.5w,-0.9h}{i33}
\fmfforce{0.5w,-1.6h}{i34}
\fmfforce{0.4w,-1h}{e31}
\fmfforce{0.1w,-1h}{e32}
\fmfforce{0.25w,-1.6h}{o31}
\fmf{plain }{i32,i31}
\fmf{plain }{i33,i34}
\fmf{dots}{e31,e32}
\fmfblob{0.5w}{o31}

\fmfforce{0w,-0.9h}{i32}
\fmfforce{0w,-1.6h}{i31}
\fmfforce{0.5w,-0.9h}{i33}
\fmfforce{0.5w,-1.6h}{i34}
\fmfforce{0.4w,-1h}{e31}
\fmfforce{0.1w,-1h}{e32}
\fmfforce{0.25w,-1.6h}{o31}
\fmf{plain }{i32,i31}
\fmf{plain }{i33,i34}
\fmf{dots}{e31,e32}
\fmfblob{0.5w}{o31}

\fmfforce{1.5w,-0.9h}{i42}
\fmfforce{1.5w,-1.6h}{i41}
\fmfforce{2w,-0.9h}{i43}
\fmfforce{2w,-1.6h}{i44}
\fmfforce{1.9w,-1h}{e41}
\fmfforce{1.6w,-1h}{e42}
\fmfforce{1.75w,-1.6h}{o41}
\fmf{plain }{i42,i41}
\fmf{plain }{i43,i44}
\fmf{dots}{e41,e42}
\fmfblob{0.5w}{o41}

\fmfforce{0.5w, -0.2}{x1}
\fmfforce{1.5w, -0.2}{x2}
\fmf{dbl_dashes}{i42,i33}
\fmf{plain}{x1,i33}
\fmf{plain}{x2,i42}
\fmfv{decor.shape=circle,decor.filled=full,
decor.size=0.07w}{i33}
\fmfv{decor.shape=circle,decor.filled=full,
decor.size=0.07w}{i42}

\fmfforce{0.5w , 1h}{lab1}
\fmfforce{1.15w , 1.1h}{lab2}
\fmfforce{0.27w , -0.39h}{lab3}
\fmfforce{1.57w , -0.39h}{lab4}
\fmflabel{$\phi_1; w  \approx 0$}{lab1}
\fmflabel{$\phi_2;w  \approx W$}{lab2}
\fmflabel{$\phi_1;w  \approx 0$}{lab3}
\fmflabel{$\phi_2;w  \approx W$}{lab4}

\fmfforce{0.5w , -0.12h}{r1}
\fmfv{decor.shape=circle,decor.filled=full,
decor.size=0.07w}{r1}
\fmfforce{1.5w , -0.12h}{r2}
\fmfv{decor.shape=circle,decor.filled=full,
decor.size=0.07w}{r2}
\fmfforce{0.6w , -0.1h}{r1bis}
\fmfforce{1.12w , -0.1h}{r2bis}
\fmflabel{$x_1$}{r1bis}
\fmflabel{$x_2$}{r2bis}

\fmfforce{0.57w,-0.9h}{i33bis}
\fmfforce{1.27w,-0.9h}{i42bis}
\fmflabel{$z$}{i33bis}
\fmflabel{$z$}{i42bis}

\end{fmfgraph*}
\end{fmffile}
\end{gathered} \hspace{3.9cm}}
\end{eqnarray}
It can be written as 
\bea \label{localZc} \nn
&& \hat Z^{c,\phi^1 \phi^2}_{W}(\lambda_1,\lambda_2 ) = -  \frac{\Delta''(W)}{\int_x \phi^1_x  \int_x \phi^2_x } \\
&&  \int_{z x_1 x_2 y_1 y_2} g_{zx_1} Z_{x_1}^{\phi^1}(\lambda_1) \frac{ \delta Z^{\phi^1}_{y_1}(\lambda_1)}{\lambda_1 \delta \phi^1_z} g_{zx_2} Z_{x_2}^{\phi^2}(\lambda_2) \frac{ \delta Z^{\phi^2}_{y_2}(\lambda_2)}{\lambda_2 \delta \phi^2_z}  \nn \\
&& + O(\epsilon^2)   \  .
\eea
We note that it is possible to obtain a more explicit formula for   avalanches measured on parallel hyperplanes, see Appendix \ref{app:ZlocHyperplane}. In the next section we focus on the first moments which already contain valuable information.

\subsection{First moments: arbitrary sources and kernels}

The first moments of $\rho^{c;\phi^1 \phi^2}_W$ are   obtained from the combination of Eqs.~(\ref{local:hatZ}), (\ref{local:hatZandZ}), (\ref{localDisc}) and (\ref{localZc}). One first needs the series expansion for $ Z^{\phi}_x(\lambda)$. It is   obtained from Eq.~(\ref{local:selfcons}) at arbitrary order in $\lambda$;  here we give it up to order $3$:
\bea
\lefteqn{Z^{\phi}_x(\lambda) = \lambda \phi_x + \lambda^2 \sigma \int_{yy'} g_{x-y} g_{x-y'} \phi_y \phi_{y'}} \\
&& +2\lambda^3 \sigma^2 \int_{yy'zz'} g_{x-y}g_{x-y'} g_{y-z}g_{y-z'} \phi_z \phi_{z'} \phi_{y'} + O(\lambda^4) \nn 
\eea 
Hence
\bea
\lefteqn{ \frac{\delta Z_x^{\phi}}{\lambda \delta \phi_u} = \delta(x-u) + 2 \lambda \sigma \int_{y}g_{x-y} g_{x-u} \phi_y  }\nn \\
&& +  2 \lambda^2 \sigma^2 \left(2  \int_{yy'z} g_{x-y}g_{x-y'} g_{y-z}g_{y-u} \phi_z  \phi_{y'}    \right. \nn \\
&& \left . + \int_{yzz'} g_{x-y}g_{x-u} g_{y-z}g_{y-z'} \phi_z \phi_{z'}  \right) +  O(\lambda^3)\ .~~~~~~~~~~~
\eea
We then obtain from Eq.~(\ref{localZc}) the local version of the exact result (\ref{S1S2exactIntro}), namely \footnote{The result (\ref{S1S2loc}) can simply be turned into an exact one if one introduces the bi-local part of the renormalized disorder correlator $\Delta_{x_2-x_1}(w_1-w_2) = m^4 \overline{[u_{x_1}(w_1) -w_1][u_{x_2}(w_2) -w_2]}$ (see also \cite{LeDoussal2008}) and proceeds as in Sec.~\ref{Subsec:FirstMom}. The result (\ref{S1S2loc}) can then be understood as the lowest-order approximation of $\Delta_{x_2-x_1}(w)$ in terms of $\Delta(w)$.} 
\bea \label{S1S2loc}
   \frac{\langle \langle S_1^{\phi^1} S_2^{\phi^2} \rangle \rangle_{\rho^{c}_W}}{\int_{x} \phi^1_x\int_{x} \phi^2_x }   &=&   - \frac{\Delta''(W) }{\int_{x} \phi^1_x\int_{x} \phi^2_x } \int_{z x_1 x_2} g_{z-x_1} g_{z-x_2}\phi^1_{x_1} \phi^2_{x_2}  \nn\\
&&+ O(\epsilon^2)\ . 
\eea
Let us also give   the result for the third-order moment, 
\bea \nn
\lefteqn{\frac{\langle \langle (S_1^{\phi^1})^2 S_2^{\phi^2} \rangle \rangle^c_{\rho^{c}_W}}{\int_{x} \phi^1_x\int_{x} \phi^2_x }  = -\frac{\Delta''(W) }{\int_{x} \phi^1_x\int_{x} \phi^2_x }\sigma \times }   \\
&&   \bigg( 4  \int_{z x_1 x_2 y_1 t_1} g_{z-x_1}  g_{z-x_2}  g_{y_1-t_1} g_{y_1-z} \phi^1_{x_1} \phi^1_{t_1} \phi^2_{x_2}  \nn \\
&& \left. +2 \int_{z x_1 x_2 t_1 t_1'} g_{z-x_1}  g_{z-x_2} g_{x_1 -t_1} g_{x_1- t_1'}  \phi^1_{t_1}  \phi^1_{t_1'} \phi^2_{x_2}    \right) \nn\\
&& + O(\epsilon^2)  \ .
\eea

\subsection{First moment: correlations between the local shock sizes for short-ranged elasticity.}

Let us now give the precise form of the first connected moment for an interface with the short-ranged elasticity (\ref{short-rangedKernel}) and for correlations between the local avalanche  sizes at two points $x_1$ and $x_2$.  We choose $\phi^1_x = \delta^d(x-x_1)$ and $\phi^2_x = \delta^d(x-x_2)$ and note $x = |x_1-x_2|$ the distance between the two points. Thus $S_1^{\phi^1} = S_{1x_1}$ and $S_2^{\phi^2} = S_{2x_2}$. We obtain
\bea \label{S1S2locSR}
 \lefteqn{\langle \langle S_{1x_1} S_{2x_2} \rangle \rangle_{\rho^{c}_W} =  - \Delta''(W) \int_{q} e^{iq(x_1-x_2)} g_q g_{-q} } \nn \\
&&= -  \Delta''(W) m^{d-4} 2^{-\frac{d}{2}-1} \pi ^{-\frac{d}{2}} (m x)^{2-\frac{d}{2}} K_{2-\frac{d}{2}}(m x)   \nn \\
&& =_{x=0} -  \Delta''(W) 2^{-d} \pi ^{-\frac{d}{2}} m^{d-4} \Gamma \left(2-\frac{d}{2}\right) \nn \\
&& \simeq_{x \gg 1/m} -\Delta''(W)  2^{-\frac{d}{2}-\frac{3}{2}} \pi ^{\frac{1}{2}-\frac{d}{2}} m^{\frac{d-5}{2}} x^{\frac{3}{2}-\frac{d}{2}} e^{-m x}  \ , ~~~~~~~\label{110}
\eea
where $K_n(x)$ denotes a modified Bessel function of the second kind. Note that integrating this  formula yields an exact result,
\be
\int_{x_1 , x_2} \langle \langle S_{1x_1} S_{2x_2} \rangle \rangle_{\rho^{c}_W} = \langle S_1 S_2 \rangle_{\rho^{c}_W} = - L^d\frac{\Delta''(W)}{m^4}  \ . 
\ee
This is equivalent to Eq.\ (\ref{S1S2exact2}), which is {\em exact}. We thus expect Eq.\ (\ref{110}) to be quite accurate even for  large values of $\epsilon$.

As expected, we observe that the amplitude of the correlations decays exponentially beyond the length $L_m=1/m$. For smaller distances they decay algebraically with an exponent that depends on the dimension:
\bea
\lefteqn{ \langle \langle S_{1x_1} S_{2 x_1+x}\rangle \rangle_{\rho^{c}_W} -\langle \langle S_{1x_1} S_{2x_1} \rangle \rangle_{\rho^{c}_W} } \nn \\
&& \simeq_{d=1}   \frac{\Delta''(W)}{8 m} x^2 + O(x^3) \nn \\
&& \simeq_{d=2} -  \frac{\Delta''(W)}{16 \pi} \Big[2 \gamma_{\rm E} -1  + 2 \log(mx/2)\Big]x^2 \nn \\
&& \simeq_{d=3} \frac{\Delta''(W)}{8 \pi} x + O(x^2) \ .
\eea
Finally, to emphasize the universal nature of Eq.~(\ref{S1S2locSR}), we note that it can be rewritten, using the notations of Sec.~\ref{subsec:Analysis} and introducing a new universal scaling function ${\cal F}^{11}_d(w ,  x)$, as
\bea
  \langle \langle S_{1x_1} S_{2x_2} \rangle \rangle_{\rho^{c}_W} &=& {\cal F}^{11}_d( \frac{W}{W_{\mu}},  m|x_1-x_2| )  \\
  {\cal F}^{11}_d(w ,  x)  &=& -  2^{-\frac{d}{2}-1} \pi ^{-\frac{d}{2}}  A_d \Delta^{* \prime \prime} (w)  x^{2-\frac{d}{2}} K_{2-\frac{d}{2}}(x)  \nn\\
 && + O(\epsilon^2)    \ .
\eea

\subsection{First moment: correlations between the local shock sizes for long-ranged elasticity.}
Let us now study the correlations between local avalanche  sizes (we choose again $\phi^1_x = \delta^d(x-x_1)$ and $\phi^2_x = \delta^d(x-x_2)$ with $|x_1-x_2|=x$) for the case of long-ranged elasticity using the kernel (\ref{long-rangedKernel}) with $\gamma = 1$. Then the result for the first connected moment is
\bea
\langle \langle S_{1x_1} S_{2x_2} \rangle \rangle_{\rho^{c}_W} & =&  - \Delta''(W) \frac{\mu ^{d-2}}{(2 \pi )^{\frac{d}{2}} } (\mu  x)^{1-\frac{d}{2}} K_{1-\frac{d}{2}}(x \mu ) \nn \\
&& =_{d=1} \frac{e^{-\mu x}}{2 \mu} \ .
\eea
As   the previous formula   for  short-ranged elasticity, this formula should be rather accurate for the experimentally relevant case of $d=1$ (in this case $\epsilon=1$). We again observe an exponential decay of the correlations beyond the length $L_\mu=1/\mu$. However, here  the correlations are constant at small distances, a signature of the long-range nature of the elasticity. As before, the universal nature of this result can be emphasized by introducing a universal scaling function ${\cal F}^{11}_{d,{\rm LR}}(w ,  y)$: 
\begin{align}
 &\langle \langle S_{1x_1} S_{2x_2} \rangle \rangle_{\rho^{c}_W} ={\cal F}^{11}_{d,{\rm LR}} \left(\frac{W}{W_{\mu}} ,  \mu |x_1-x_2| \right)  \\
 & {\cal F}^{11}_d(w ,  x)  = -  (2 \pi )^{-\frac{d}{2}}  A_d \Delta^{* \prime \prime} (w)  x^{1-\frac{d}{2}} K_{1-\frac{d}{2}}(x)  + O(\epsilon^2) \nn  \ ,
\end{align}
where we used the same notations as in Sec.~\ref{subsec:Analysis}.

\section{Measurement of correlations in simulations of $d=0$ toy models.} \label{Sec:Numerics}
\subsection{Models and goals}

In this section we compare our results with numerical simulations of toy models of a particle in a discrete random potential. The position of the particle can only take integer values $u \in \mathbb{N}$ and its Hamiltonian is
\be 
{\cal H}_V[u;w]=V(u) + \frac{1}{2} m^2(u-w)^2 \ ,
\ee 
where $V$ is a random potential. We consider two distributions for the random potential mimicking the two non-periodic static universality classes of interfaces models:

\paragraph*{RB model:} The first model is a toy model for the Random-Bond universality class with short-ranged correlated disorder where the random potentials $V(i)$ at each site $i \in \mathbb{N}$, are chosen as independent, centered and normalized Gaussian random variables. 

\paragraph*{RF model:} The second model is a toy model for the Random-Field universality class where   $V(0)=0$ and for $i \geq 1$, $V(i) = -\sum_{j=1}^{i} F(j)$; the random forces $F(i)$ at each site $i \in \mathbb{N}$ are chosen as independent, centered and normalized random variables. Thus $V(i)$ is a random walk with Gaussian increments.

In the RB model we choose the mass as $m_{\rm RB}  = 0.01$ and in the RF model as $m_{\rm RF}  = 0.02$. With these parameters, the probability $\rho_0$ to trigger a shock when moving $w \to w+1$ is   $\rho_0^{\rm RF} = (6.959 \pm 0.001)\times 10^{-3} $ and  $\rho_0^{\rm RB} = (9.471 \pm 0.001) \times 10^{-3} $. These small
values of the masses ensure that the models efficiently approximate our continuum model in $d=0$, and that the particle optimizes its energy over a large number of random variables. We perform averages over $10$ simulations of environments of size $N = 5 \times 10^8$ sites. We obtain excellent statistics for various observables studied in this work, including $\rho_2(W)$, $\Delta(W)$ measured using Eq.~(\ref{DeltaDef}), $\langle S_1 S_2 \rangle_{\rho_W}$ and $\langle S_1^2 S_2 \rangle_{\rho_W}$.

Let us emphasize   that these simulations are more a proof of principle to motivate simulations on higher dimensional models and measurements in experiments,  than a full test of   the results obtained in this article. 
This said, our simulations  allow us to verify the exact result (\ref{S1S2exact})  to a very high accuracy.
Second, although $d=0$ is at a large value of $\epsilon$ in  the $d=4-\epsilon$ expansion,  the FRG equation and the associated fixed-point functions for random-field disorder are known to behave quite similarly \cite{LeDoussalWiese2008a,LeDoussalWieseChauve2003}. For random-bond disorder we expect less universality since $\Delta(u)$ is non-universal in $d=0$;
nevertherless the relations between the correlation and $\Delta(u)$ are
interesting to investigate, in particular the sign of the correlations.

\subsection{Numerical Results: RB model}

Using the definition (\ref{DeltaDef}) we measure the renormalized disorder correlator. The result is shown in Fig.~\ref{fig:RBDelta}. Using an interpolation of the result with a polynomial of   degree   $10$, we obtain a smooth version that is later used to compute its second   derivative $\Delta''(u)$ which appears in our analysis as the central object controlling the amplitude of the correlations. Some measured properties are:   $\Delta(0) \approx 3.34 \times 10^{-5}$, $\Delta''(0) \approx 6.78 \times 10^{-9}$;   $\Delta(76.2) \approx 0$, $\Delta''(215) \approx 0 $; the position of the minimum and the value at the minimum: $\Delta(148 .2) \approx -7.3 \times 10^{-6}$, $\Delta''(274,4) \approx -5.1 \times 10^{-10}$. This is compared with the measurement of $\langle S_1 S_2 \rangle_{\rho_W}$ using the exact result (\ref{S1S2exact}), see Fig.~\ref{fig:RBexact}. We obtain a perfect agreement.

From a qualitative perspective, we note the following:

(i) We observe the predicted crossover from anti-correlated shocks at small distances $(W <215)$ to positively correlated shocks at large distances. 

(ii) The correlations are far from being negligible: by definition $\frac{ \langle S_1S_2 \rangle_{\rho_W^c} }{\langle S \rangle_{\rho^2}}>-1$, while   we observe $\frac{ \langle S_1S_2 \rangle_{\rho_{W \approx 0}^c} }{\langle S \rangle_{\rho}^2} \approx -0.6$, an   indication that the shocks in this toy model are  strongly correlated. 

\begin{figure}
\centerline{\includegraphics[width=8.5cm]{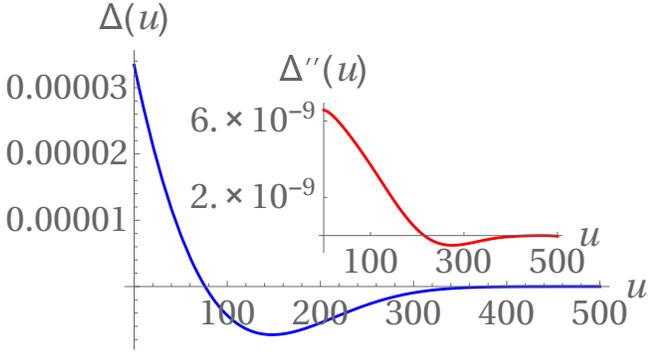}} 
\caption{Renormalized disorder $\Delta(u)$ measured in the $d=0$ RB toy model. Inset: its second derivative $\Delta''(u)$, computed using a numerical fit of the measured $\Delta(u)$.}
\label{fig:RBDelta}
\end{figure}

\begin{figure}
\centerline{\includegraphics[width=8.5cm]{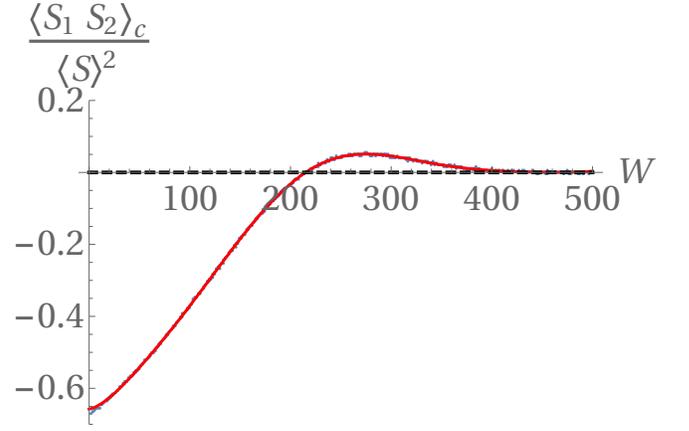}} 
\caption{Comparison between the measurement of the normalized moment $\frac{ \langle S_1S_2 \rangle_{\rho_W^c} }{\langle S \rangle_{\rho}^2}$ (blue dots) and the prediction from the exact result (\ref{S1S2exact}) using the measurement of $\Delta(u)$ (red curve) in the RB toy model. The agreement is perfect as expected.}
\label{fig:RBexact}
\end{figure}

We now check the predictions obtained using the $\epsilon$ expansion. We first measure $\rho_2(W)$ and compare it with the result (\ref{resRho2}), see Fig.~\ref{fig:RBrho}. We obtain a surprisingly good agreement between the two curves, considering that   $\epsilon=4$. We also measure $\langle S_1^2 S_2 \rangle_{\rho_W}$ and compare it with the result (\ref{resS12S2}), see Fig.~\ref{fig:RB21}. Here the discrepancy is large for   smaller
values of $W$, 
a fact that can be anticipated since our result predicts  $\frac{ \langle S_1^2S_2 \rangle_{\rho_W^c} }{\langle S^2 \rangle_{\rho} \langle S \rangle_{\rho}}<-1$ at small $W$, which is   unphysical. This discrepancy keeps increasing with higher-order moments. However the sign of the correlation, and its
value for large $W$ is quite well predicted. 

\begin{figure}
\centerline{\includegraphics[width=8.5cm]{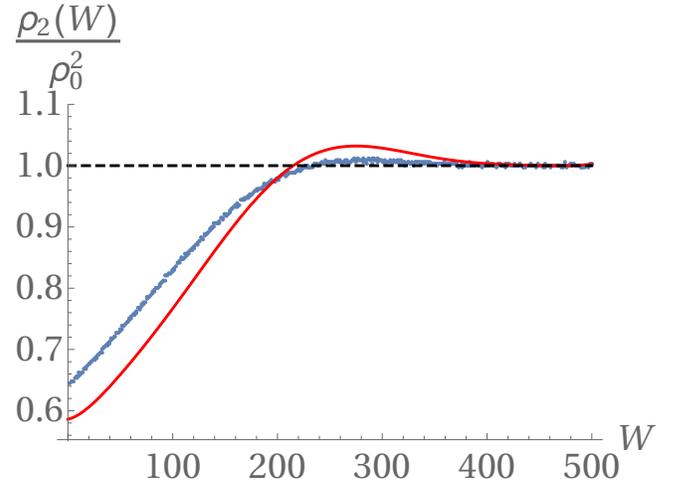}} 
\caption{Comparison between the measurement of $\rho_2(W)$ (blue dots) and the prediction from the $O(\epsilon)$ result (\ref{resRho2}) using the measurement of $\Delta(u)$ (red curve) in the RB toy model. We obtain a surprisingly good agreement.}
\label{fig:RBrho}
\end{figure}

\begin{figure}
\centerline{\includegraphics[width=8.5cm]{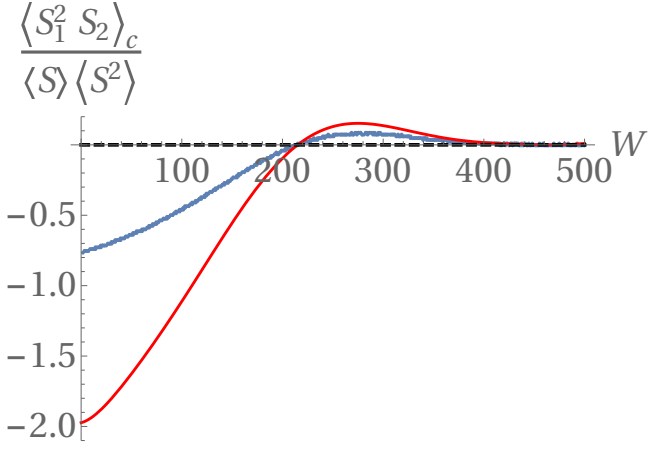}} 
\caption{Comparison between the measurement of the normalized moment $\frac{ \langle S_1^2S_2 \rangle_{\rho_W^c} }{\langle S^2 \rangle_{\rho} \langle S \rangle_{\rho}}$ (blue dots) and the prediction from the exact result (\ref{S1S2exact}) using the measurement of $\Delta(u)$ (red curve) in the RB toy model.}
\label{fig:RB21}
\end{figure}

\subsection{Numerical Results: RF model}
In Figs.~\ref{fig:RFDelta} to \ref{fig:RF21} we show the corresponding results for the  RF toy model. They are   similar except that as predicted in this type of model the shocks are {\em always anti-correlated}.
The value at the origin of the renormalized disorder correlator and of its second derivative are measured as $\Delta(0) \approx 3.4 \times 10^{-3}$, $\Delta''(0) \approx 9.4 \times 10^{-8}$. Once again we observe that these  correlations are  large, $\frac{ \langle S_1S_2 \rangle_{\rho_{W\approx 0}^c} }{\langle S \rangle_{\rho}^2} \approx -0.6$. We obtain a perfect agreement for the exact result $\langle S_1S_2 \rangle_{\rho_{W}}$, see Fig.~\ref{fig:RFexact}.
The agreement for 
 the $O(\epsilon)$ result for $\rho_2(W)$ (\ref{resRho2})  is    surprisingly good    (see Fig.~\ref{fig:RFrho}), whereas the $O(\epsilon)$ approximation breaks down for higher   moments at small $W$ such as $\langle S_1^2S_2 \rangle_{\rho_W^c}$, see Fig.~\ref{fig:RF21}.

\begin{figure}
\centerline{\includegraphics[width=8.5cm]{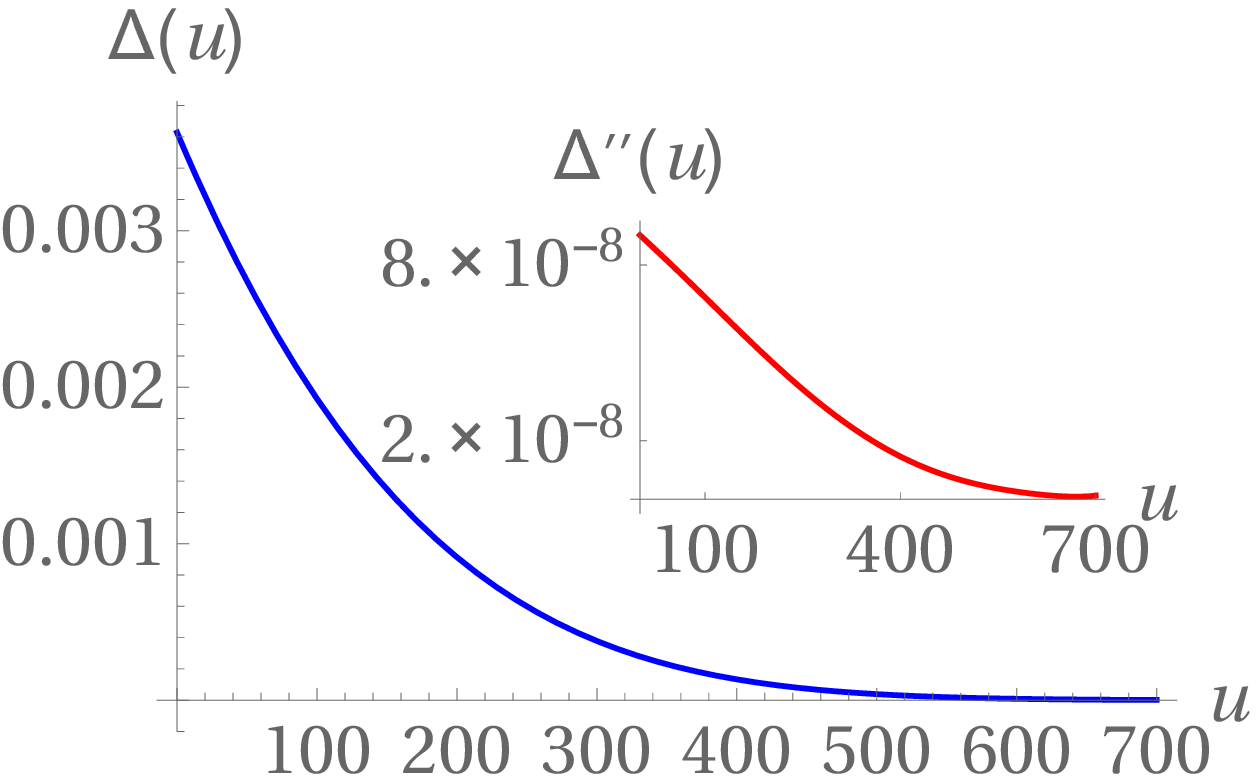}} 
\caption{Renormalized disorder $\Delta(u)$ measured in the $d=0$ RF toy model. Inset: its second derivative $\Delta''(u)$, computed using a numerical fit of the measured $\Delta(u)$.}
\label{fig:RFDelta}
\end{figure}

\begin{figure}
\centerline{\includegraphics[width=8.5cm]{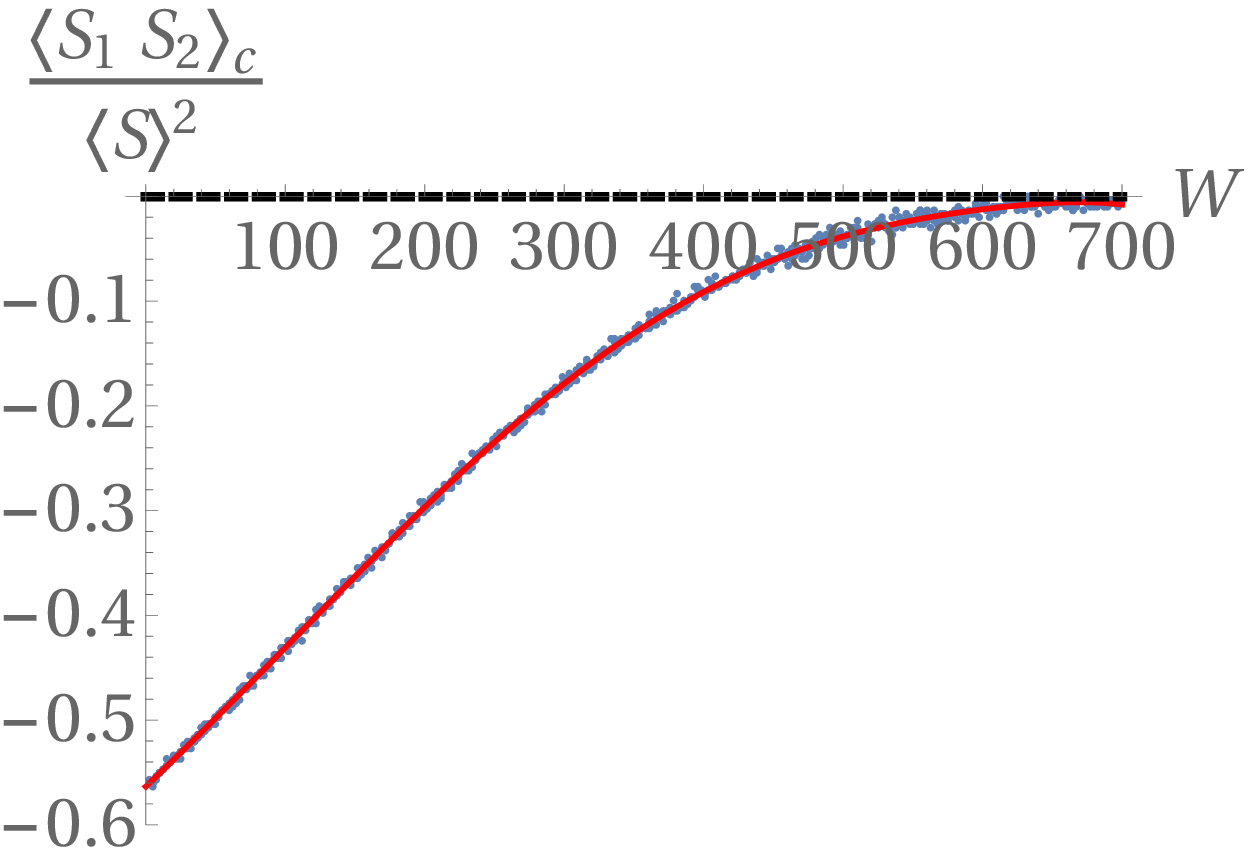}} 
\caption{Comparison between the measurement of the normalized moment $\frac{ \langle S_1S_2 \rangle_{\rho_W^c} }{\langle S \rangle_{\rho}^2}$ (blue dots) and the prediction from the exact result (\ref{S1S2exact}) using the measurement of $\Delta(u)$ (red curve) in the RF toy model. The agreement is perfect as expected.}
\label{fig:RFexact}
\end{figure}

\begin{figure}
\centerline{\includegraphics[width=8.5cm]{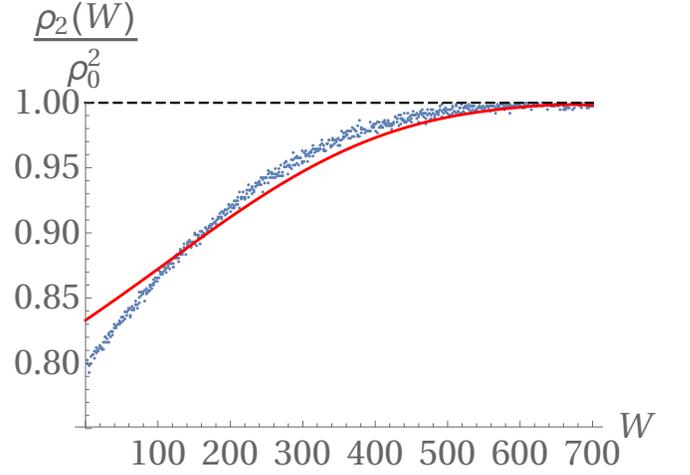}} 
\caption{Comparison between the measurement of $\rho_2(W)$ (blue dots) and the prediction from the $O(\epsilon)$ result (\ref{resRho2}) using the measurement of $\Delta(u)$ (red curve) in the RF toy model. The agreement is surprisingly good.}
\label{fig:RFrho}
\end{figure}

\begin{figure}
\centerline{\includegraphics[width=8.5cm]{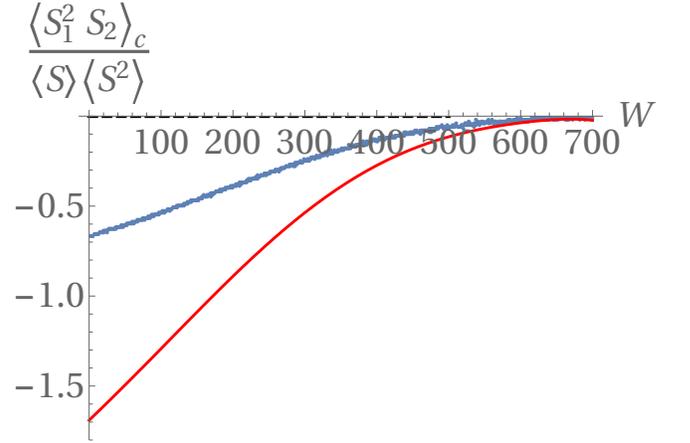}} 
\caption{Comparison between the measurement of the normalized moment $\frac{ \langle S_1^2S_2 \rangle_{\rho_W^c} }{\langle S^2 \rangle_{\rho} \langle S \rangle_{\rho}}$ (blue dots) and the prediction from the $O(\epsilon)$ result (\ref{S1S2exact}) using the measurement of $\Delta(u)$ (red curve) in the RF toy model.}
\label{fig:RF21}
\end{figure}

\section{Conclusion}

In this paper we shed light on the fact that, for realistic models of elastic interfaces in a random medium below their upper critical dimension, correlations between (static) avalanches should always be expected. To do so we have studied the correlations between the size and location of shocks in the ground state of elastic interfaces in a random potential. We found the exact relation (\ref{S1S2exactIntro}) for the first connected moment that   characterizes these correlations in terms of the renormalized disorder correlator, a universal quantity at the center of the FRG treatment of disordered elastic systems. Beyond the first cumulant, higher-order moments (\ref{Zcres}), (\ref{momentRes}) and the full joint density of shocks (\ref{resRho}) were computed using the FRG at first non-trivial order in the $\epsilon$ expansion. The local structure of these correlations was made precise through a study of local shock  sizes. The qualitative phenomenology associated with these correlations clearly distinguishes between the Random-Bond and Random-Field universality classes. This was   highlighted through a numerical simulation of $d=0$ toy models.

We expect our results to broadly apply to models in the universality class of the statics of disordered elastic systems. Concerning the dynamics, and   avalanches at the depinning transition of elastic interfaces, we   expect our results to be equivalently {\em applicable and accurate}. The derivation of the exact relation (\ref{S1S2exactIntro}) can  easily be  adapted to the dynamics by considering the quasi-static steady-state process of the position field of the interface instead of the position of its ground-state as was done in Ref.~\cite{LeDoussalWiese2012a}. For the results at the improved tree level, it is expected that both theories are equivalent for those observables \cite{LeDoussalWiese2012a}. The most important difference is that in the dynamics the Random-Bond universality class is unstable, and thus the observed correlations should always be of the Random-Field type (at least as long as the microscopic  disorder is short-ranged).

For physical systems where the usual model of elastic interfaces is accurate, our results give a precise description of the correlations. Even if additional mechanisms generating correlations are present, such as in earthquake problems, correlations due to the short-ranged nature of the disorder as described in this work should be included in order to gain a quantitative understanding of the correlations due to these additional mechanisms.

\begin{acknowledgments}
We warmly thank Alexander Dobrinevski for numerous discussions. We acknowledge hospitality from the KITP in Santa Barbara where part of this work was conducted. This research was supported in part by the National Science Foundation under Grant No. NSF PHY11-25915. We acknowledge support from
PSL grant ANR-10-IDEX-0001-02-PSL. 
\end{acknowledgments}

\appendix

\section{Proof of the identity on generating functions} \label{app:Gene}

As in the case of one shock (Appendix A of \cite{LeDoussalWiese2008c}), the important identity is
\bea
&& (\partial_{\delta} + \lambda L^d) e^{\lambda L^d (u(w+\delta)-w - \delta)} = \nn \\
&&  \sum_i (e^{\lambda S_i}-1) e^{\lambda L^d[u(w_i^-)-w- \delta]} \delta(w +\delta-w_i)~~~~~~~~~~~~~~~~~
\eea
By definition $u(w_i^-) = L^{-d} \sum_{j < i} S_j$. Let us consider
\begin{eqnarray}
\lefteqn{ G_{w_1,w_2}(\delta_1 , \delta_2) = (\partial_{\delta_1} + \lambda_1 L^d)   (\partial_{\delta_2} + \lambda_2 L^d) \times }\nn \\
&&  e^{\lambda_1 L^d [u(w_1+\delta_1)-u(w_1)-\delta_1] +\lambda_2 L^d [u(w_2+\delta_2)-u(w_2)-\delta_2]} \nn \\
\lefteqn{ =  \sum_{ij}( e^{\lambda_1 S_i} - 1 ) ( e^{\lambda_2 S_j} - 1 ) e^{ \lambda_1 L^d [ u(w_i^-)-u(w_1) -\delta_1]} \times } \nn \\
&& e^{\lambda_2 L^d [u(w_j^-)-u(w_2) -\delta_2]} \delta(w_1+\delta_1-w_i) \delta(w_2+\delta_2-w_j) \nn\\
\end{eqnarray} 
Taking advantage of the Dirac $\delta$-function, we can replace the $u(w_1)$ inside the exponential by $u(w_i - \delta_1)$ which unambiguously gives $u(w_i^-)$ when one takes the limit of $\delta_1 \to 0^+$. We thus obtain
\bea 
\lefteqn{\lim_{\delta_1, \delta_2 \to 0^+} G_{w_1,w_2}(\delta_1 , \delta_2) = }\nn \\
&& \sum_{ij} (e^{\lambda_1 S_i} -1 ) (e^{\lambda_2 S_j} -1 ) \delta(w_1 - w_i)\delta(w_2 -w_j). ~~~~~~~ 
\eea
Taking the average over disorder, we obtain by definition of $Z_{w_2-w_1}(\lambda_1 , \lambda_2) $
\be 
Z_{w_2-w_1}(\lambda_1 , \lambda_2) = \lim_{\delta_1, \delta_2 \to 0^+} L^{-2d} \overline{G_{w_1,w_2}(\delta_1 , \delta_2) }\ .
\ee 
On the other hand, developing $(\partial_{\delta_1} + \lambda_1 L^d)   (\partial_{\delta_2} + \lambda_2 L^d) = \partial_{\delta_1}\partial_{\delta_2} +L^d \lambda_1 \partial_{\delta_2}  +L^d \lambda_2 \partial_{\delta_1}+L^{2d} \lambda_1 \lambda_2$ in the expression of $G_{w_1,w_2}(\delta_1 , \delta_2)$ one   arrives at Eqs.~(\ref{Z3}) and (\ref{Z2shocks2}).

\section{Diagrammatic rules for two-shocks moments} \label{app:DiagramRule}

In this appendix we explain the rules (i), (ii) and (iii) that were stated and used in Sec.~\ref{Sec:Global} to obtain diagrammatically the result (\ref{Zcres}). \\

These rules come from the fact that in the ${\cal K}$ operation each external leg produces an additional factor of $\delta_1$ (for the $n$ legs at $w_1 , \dots w_n \approx 0$) or $\delta_2$ (for the $m$ legs at $w_{n+1} , \dots w_{n+m}  \approx W$), thus tend to be of higher order in $\delta_1$ and $\delta_2$. However, from the study of the one-shock case (see Section V.C of \cite{LeDoussalWiese2008c}), we   know the general mechanism to escape this apparent trivialization and to allow that each part of the diagram that connects only coinciding points together brings a single $\delta_i$. In this case, starting from the top of a diagram the $\delta_i$ attached to an external leg can be brought to the bottom of the diagram as long as the disorder vertex encountered along the way leads to a $\Delta'(0^+)$ when taking the limit of coinciding points. In such diagrams each vertex linking coinciding points must have two up-going propagators and one entering from below (effectively corresponding to the $\Delta'(0^+)$ cubic vertex of the BFM \cite{LeDoussalWiese2012a}), except for the vertex at the bottom of the diagram which has only two up-going propagators (see Section V.D. of \cite{LeDoussalWiese2008c}). This last vertex is the one carrying the remaining factor of $\delta_1$: being differentiated in the end it also leads to an additional factor of $\Delta'(0^+)$. This explains why the disorder only enters as $\Delta'(0^+)$ in the one-shock improved-tree-theory result (\ref{1shockResult}). The rule (iii) stated above is a generalization of that property.

In the two-shock case the same mechanism occurs and rule (i)  is obvious: a diagram cannot have more than two sets of points separated by a double-dashed line (one around $w  \approx 0$ and one around $w  \approx W$) since each set contributes a factor of $\delta_i$. For example,  in the last diagram of Eq.~(\ref{thirdMomDiagram}), each leg is such a set of points, and the diagram is $O(\delta_1^2 \delta_2)$. To explain rule (ii), let us consider one endpoint of a double-dashed line and distinguish three cases.
First, if there is no propagator entering from below this point, such as the points at $w \approx W$ in the first and second diagrams of Eq.~(\ref{thirdMomDiagram}) and the two points in Eq.~(\ref{SecondMomDiagram}), then the $\delta_i$ originating from the set of connected points above it end at this vertex, and the vertex is differentiated during the ${\cal K}$ operation. Second, if there is a propagator entering from below that point, such as the point at $w_1  \approx 0$ in the first diagram of Eq.~(\ref{thirdMomDiagram}), then the $\delta_i$ originating from above the vertex continues  downward the diagram without modifying the vertex. Third, if there is more than one propagator entering from below the point then the diagram will necessarily 
be of higher order in $\delta_i$. Combining these three cases, one concludes that the double-dashed-line vertex necessarily corresponds to a $\Delta''(W)$.

\section{A derivation from the Carraro-Duchon formula} \label{app:C-D}

Let us recall the results obtained in Ref.~\cite{LeDoussalWiese2011b}, generalizing to arbitrary dimension the result from Ref.~\cite{CarraroDuchon1998}.  Consider
\begin{equation} \label{defZcarraro}
e^{ L^d \hat{\Z}_t \{ \omega_i , w_i \}}:= \overline{e^{\frac{-L^d}{t}\sum_{i=1}^p \omega_i [ u(w_i) - w_i]} } \ ,
\end{equation}
where   $t := \frac{1}{m^2}$. Then, in the improved-tree theory, $\hat{\Z}$ solves the differential equation 
\bea \label{C-D:eq1}
&& \partial_t \hat{\Z}_t \{ \omega_i , w_i \} = - \sum_{i=1}^p \frac{\partial}{\partial \omega_i}  \hat{\Z}_t \{ \omega_i , w_i \}  \frac{\partial}{\partial w_i}  \hat{\Z}_t \{ \omega_i , w_i \} \nn \\
&& \hat{\Z}_{t=0} \{ \omega_i , w_i \} = \frac{1}{2} \sum_{i,j=1}^p \omega_i \omega_j \Delta( w_i - w_j)\ .
\eea
It further satisfies the STS symmetry relation, 
\bea \label{C-D:STS}
&& \hat{\Z}_t \{ \omega_i , w_i + \delta w \} = \hat{\Z}_t \{ \omega_i , w_i \} \nn \\
&& \sum_i \frac{\partial}{\partial w_i}  \hat{\Z}_t \{ \omega_i , w_i \} = 0 \ .
\eea
In order to extract the needed information for the two-shock statistics we  choose   $p=4$ and the quadruplets 
$(\omega_1,\omega_2,\omega_3,\omega_4)=(-\omega_1- \tilde \omega, \omega_1,-\omega_2+\tilde \omega, \omega_2)$ and 
$(w_1,w_2,w_3,w_4)=(0,\delta_1,W,W+\delta_2)$. We then consider  (with a slight abuse of notations)
\bea
&& \tilde{\Z}_t( \omega_1 ,\delta_1 , \tilde{\omega} , W , \omega_2 , \delta_2)  \\
&&  = \hat{\Z}_t( -\omega_1 - \tilde \omega , 0 , \omega_1, \delta_1 , -\omega_2 +\tilde{\omega} , W , \omega_2 , W + \delta_2)  \nn
\eea
Because of the STS the $p=4$ function $\tilde{\Z}_t$ depends only on
six variables (and not eight) and satisfies a closed equation. Indeed, 
using Eqs.~(\ref{C-D:eq1}) and (\ref{C-D:STS}), one proves that $\tilde{\Z}_t$ satisfies the following evolution equation
\begin{equation} \label{C-D:eq2}
\partial_t \tilde{\Z}_t = - \left( \frac{\partial}{\partial \omega_1} \tilde{\Z}_t  \frac{\partial}{\partial \delta_1} \tilde{\Z}_t  + \frac{\partial}{\partial \tilde{\omega}} \tilde{\Z}_t  \frac{\partial}{\partial W} \tilde{\Z}_t  + \frac{\partial}{\partial \omega_2} \tilde{\Z}_t  \frac{\partial}{\partial \delta_2} \tilde{\Z}_t  \right)
\end{equation} 
We are only   interested in a perturbative resolution. Define the expansion
\begin{equation} \label{C-D:expansion}
\tilde{\Z}_t = \sum_{m n p} z_{mn}^p (t,  \omega_1 , \omega_2 , W) \delta_1^m \delta_2^n \tilde w^p \ .
\end{equation}
Indeed, this is sufficient to retrieve the generating function $\hat{Z}_W(\lambda_1, \lambda_2) =\hat{Z}^{{\rm disc}}_W(\lambda_1, \lambda_2)  + \hat{Z}^c_W(\lambda_1, \lambda_2)$ as (compare with the small-$\delta_i$ expansion of (\ref{defZcarraro}) and (\ref{Z2shocks2}))
\begin{eqnarray}
&& \hat{Z}^{{\rm disc}}_W(\lambda_1 , \lambda_2) = z_{10}^0( \omega_1  ,\omega_2   , W ) z_{01}^0( \omega_1 ,\omega_2  , W ) \nonumber \\
&& \hat{Z}^c_W(\lambda_1 , \lambda_2) = L^{-d}z_{11}^0( \omega_1   ,\omega_2 , W)\ .
\end{eqnarray}
On the right-hand side the arguments are $\omega_1 = -t \lambda_1$ and $\omega_2 = -t \lambda_2$. Inserting the expansion (\ref{C-D:expansion}) inside Eq.~(\ref{C-D:eq2}), we obtain the initial conditions:
\begin{eqnarray} \label{appCarraCI}
z_{00}^0(t=0 , \omega_1 , \omega_2 , W) &  =  & 0
\nonumber \\
z_{10}^0(t=0 , \omega_1 , \omega_2 , W) &  =  & - \Delta'(0^+) \omega_1^2 \nonumber \\
z_{01}^0(t=0 , \omega_1 , \omega_2 , W) &  =  & - \Delta'(0^+) \omega_2^2 \nonumber \\
z_{11}^0(t=0 , \omega_1 , \omega_2 , W) &  =  & - \Delta''(W) \omega_1 \omega_2 \nonumber \\
z_{00}^1(t=0 , \omega_1 , \omega_2 , W) & = & 0 \label{carr-init} \ .
\end{eqnarray}
Obviously we have $z_{00}^0(t , \omega_1 , \omega_2 , W)  = 0 $, $\forall t$. We also obtain the evolution equation: 
\begin{eqnarray} \label{appCarraSyst}
\partial_t z_{10}^0 &=&  - \Big(\frac{\partial}{\partial \omega_1} z_{10}^0 \Big)z_{10}^0 - \Big(\frac{\partial}{\partial \omega_2} z_{10}^0 \Big)z_{01}^0  - z_{00}^1  \frac{\partial}{\partial W} z_{10}^0  \nonumber \\
\partial_t z_{01}^0 &=&  - \Big(\frac{\partial}{\partial \omega_1} z_{01}^0 \Big)z_{10}^0 - \Big(\frac{\partial}{\partial \omega_2} z_{01}^0 \Big)z_{01}^0 - z_{00}^1  \frac{\partial}{\partial W} z_{01}^0  \nonumber \\
\partial_t z_{11}^0 &=&  - \Big(\frac{\partial}{\partial \omega_1} z_{10}^0 \Big)z_{11}^0  - \Big(\frac{\partial}{\partial \omega_1} z_{01}^0 \Big) 2 z_{20}^0  - \Big(\frac{\partial}{\partial \omega_1} z_{11}^0 \Big) z_{10}^0 \nn \\
&& - \Big(\frac{\partial}{\partial \omega_2} z_{10}^0 \Big) 2 z_{02}^0  - \Big(\frac{\partial}{\partial \omega_2} z_{01}^0 \Big)z_{11}^0  - \Big(\frac{\partial}{\partial \omega_2} z_{11}^0 \Big)z_{01}^0  \nonumber \\
&& -  z_{00}^1  \frac{\partial}{\partial W} z_{11}^0-  z_{10}^1  \frac{\partial}{\partial W} z_{01}^0-  z_{01}^1  \frac{\partial}{\partial W} z_{10}^0 \nonumber \\
\partial_t z_{00}^1 &=&   -  \Big(\frac{\partial}{\partial \omega_1} z_{00}^1 \Big)z_{10}^0 -  \Big(\frac{\partial}{\partial \omega_2} z_{00}^1 \Big)z_{01}^0 - z_{00}^1 \frac{ \partial}{\partial W} z_{00}^1 \nn \\
\end{eqnarray}
As a consequence of the initial conditions (\ref{appCarraCI}),  one can look for a solution of Eq.~(\ref{appCarraSyst}) such that
\begin{equation}
\frac{\partial}{\partial \omega_2} z_{10}^0 = \frac{\partial}{\partial \omega_1} z_{01}^0  = \frac{\partial}{\partial W} z_{10}^0 = \frac{\partial}{\partial W} z_{01}^0  = z_{00}^1 = 0 \ . \label{simp1} 
\end{equation}
Each term has an interpretation in the notations of the main text. $z_{10}^0$ corresponds to $\hat Z(\lambda_1)$ and $z_{01}^0$ corresponds to $\hat Z(\lambda_2)$, which in the present notations reads (see Eqs.~(\ref{Z1shock1}) and (\ref{1shockResult}) and   recall $S_m = \sigma/m^4= \sigma t^2$)
\be 
     z_{10}^{0}(\omega_i) =  \hat Z(\lambda_i) = \frac{1 + 2 \sigma \omega_i t - \sqrt{ 1 + 4 \sigma \omega_i t} }{ 2 \sigma t^2}  \ .
\ee 
This is the solution of Eq.~(\ref{appCarraSyst}) using Eq.~(\ref{simp1}).
Note that  $ z_{00}^1 = 0 $ can be seen as the signature that diagrams contributing to the avalanche at $w=0$ and at $w=W$ can be linked only by one vertex $\Delta''(W)$, as observed in the diagrammatics, see Eq.~(\ref{ZWdiagram}).  This     is   already present in the initial condition (\ref{appCarraCI}). The equation for $z_{11}^0$ becomes 
\bea
\partial_t z_{11}^0 &=&  - \Big(\frac{\partial}{\partial \omega_1} z_{10}^0 \Big)z_{11}^0 - \Big(\frac{\partial}{\partial \omega_1} z_{11}^0\Big ) z_{10}^0 \nn \\
&& - \Big(\frac{\partial}{\partial \omega_2} z_{01}^0 \Big)z_{11}^0  - \Big(\frac{\partial}{\partial \omega_2} z_{11}^0 \Big)z_{01}^0 \ . ~~~~~~~~~
\eea
One can check that the result (\ref{Zcres}) obtained diagrammatically in the main text, and which in the
present notations reads
\be 
   z^{0}_{11} =
 - \frac{\Delta''(W)}{4 \sigma^2 t^2} \frac{1 - \sqrt{ 1 + 4 \sigma \omega_1 t}}{\sqrt{ 1 + 4 \sigma \omega_1 t}} \frac{1 - \sqrt{ 1 + 4 \sigma \omega_2 t}}{\sqrt{ 1 + 4 \sigma \omega_2 t}} \nn \ ,
\ee 
solves this equation with the initial condition (\ref{carr-init}). This demonstrates the
equivalence of the two methods and results.

\section{Saddle-point calculation for the local structure.}\label{app:Algebraic}

\subsection{Algebraic derivation of Eq.~(\ref{localZc})}

In this appendix we prove formula (\ref{localZc}) ``from first principles'' using a saddle-point calculation on the improved action (\ref{Action3}). This computation is similar to the one presented in Ref.~\cite{LeDoussalWiese2011b} for the calculation of the one-shock density. Here the observable of interest is 
\bea \label{app:alg1}
\lefteqn{ \hat Z^{\phi^1\phi^2}_{W}(\lambda_1,\lambda_2 ) = \frac{1}{\int_x \phi^1_x\int_x \phi^2_x} \lim_{\delta_1, \delta_2 \to 0^+}   \partial_{\delta_1 , \delta_2} G_{W} (\delta_1,\delta_2)} \nn \\
\lefteqn{  G_{W} (\delta_1,\delta_2)  =}  \\
&&  \overline{ e^{\int_x \phi^1_x \lambda_1(\hat u_x(w_1+\delta_1) - \hat u_x(w_1)  )} e^{ \int_x \phi^2_x  \lambda_2(\hat u_x(w_2+\delta_2) - \hat u_x(w_2)  )}} , \nn
\eea
where $w_2= w_1+W$. This observable can be expressed using the improved action $\Gamma[u]$ of the replicated field theory (\ref{Action3}) with $i=1, \dots,4$ sets of $a=1, \dots,n$ replicated position fields $\tilde{u}_{ax}^{i}$ feeling a parabolic well at position $\tilde{w}_i$ with $\tilde{w}_1 = w_1$, $\tilde w_2 = w_1 + \delta_1$, $\tilde{w}_3 = w_1+W$, $\tilde w_4= w_1 +W+ \delta_2$:
\be
G_{W} (\delta_1,\delta_2)=\int {\cal D}[u] e^{\int_x  
 \sum_{i=1}^{4} \nu_i  \psi^i_x  (u_{1x}^{i} - \tilde{w}_i)  - \Gamma[u]}
\ee 
Here and for the rest of this appendix, the $n \to 0$ limit is implicit.
To compute the disorder average we have singled out replica $a=1$. In order to write the formulas in a compact form, we   introduced new variables $\nu_2 = \lambda_1$, $\nu_1 = - \lambda_1$, $\nu_4 = \lambda_2$, $\nu_3 = - \lambda_2$, $\psi^1_x = \psi^2_x  = \phi^1_x $, $\psi^3_x = \psi^4_x  = \phi^2_x $. At the improved tree level, the functional integral is evaluated through a saddle-point calculation as
\be  \label{Gw1} 
G_{W} (\delta_1,\delta_2)= e^{\int_x  \sum_{i=1}^{4} \nu_i  \psi^i_x  (u_{1x}^{i} - \tilde{w}_i) - \Gamma[u] } \ ,
\ee 
where the position fields $u_{ax}^{i}$ solve the saddle-point equation
\begin{equation} \label{app:sdp1}
\int_{x'} g^{-1}_{xx'} (u^i_{ax'}-\tilde w_i) -\frac{1}{T} \sum_{cj} R'(u^i_{ax}- u^j_{cx}) = T \nu_i \psi^i_x \delta_{a1} \ .
\end{equation}
We are interested in the solution of Eq.~(\ref{app:sdp1}) in the $T \to 0$ limit. As in Ref.~\cite{LeDoussalWiese2011b}, we look for a solution that isolates the first replica $(a=1)$ in each set $(i=1,\dots,4)$ of position fields as
\begin{equation}\label{app:sdpans}
u^i_{ax} = u^i_{x} - (1 - \delta_{a1}) T U^i_x   .
\end{equation}
Inserting the Ansatz (\ref{app:sdpans}) into (\ref{app:sdp1}) leads to
\begin{eqnarray} \label{app:sdp2}
&& \int_{x'} g_{xx'}^{-1} (u^i_{1x'} -  \tilde w_i) + \sum_j R''(u^i_{1x}-u^j_{1x}) U^j_{x} = 0 \nonumber \\
&& \int_{x'} g_{xx'}^{-1}  U^i_{ x'} + \sum_{j\neq i} R'''(u^i_{1x} - u^j_{1x}) U^i_{x} U^j_{x} = \nu_i \psi^i_{x} \ . \nonumber
\end{eqnarray}
Being ultimately   interested in the computation of (\ref{app:alg1}), we solve this equation in an expansion in $\delta_1$ and $\delta_2$ as
\bea
&& u^1_{x} = \frac{\tilde w_1+\tilde w_2}{2} - u^{11}_{x} \delta_1 + u^{12}_x \delta_2 \nn \\
&& u^2_{x} = \frac{\tilde w_1+\tilde w_2}{2} + u^{21}_x \delta_1 + u^{22}_x \delta_2 \nn \\
&& u^3_{x} = \frac{\tilde w_3+\tilde w_4}{2} + u^{31}_x \delta_1 - u^{32}_x \delta_2   \nn \\
&& u^4_{x} = \frac{\tilde w_3+\tilde w_4}{2} + u^{41}_x \delta_1 + u^{42}_x \delta_2 \nn \\
&& U^i_{x} = U^{i0}_x + U^{i1}_x \delta_1 + U^{i2}_x \delta_2 .
\eea
Using now the definition (\ref{app:alg1}) we need to perform the following derivatives of
(\ref{Gw1}), $\partial_{\delta_1} \partial_{\delta_2} = \partial_{\tilde w_2} \partial_{\tilde w_4}$.
Since the fields $u^i_{ax}$ are evaluated at the saddle point, we can differentiate only
with respect to the explicit dependence in the $\tilde w_i$. Using the form (\ref{Action3}) for
$\Gamma[u]$, these derivatives can   be calculated by repeating the identity
\be 
\partial_{\tilde w_i} G_{W} = \left( - \nu_i \int_x \psi^i_x + \frac{1}{T} \sum_{a i} \int_{xx'} 
g_{xx'}^{-1} (u^i_{a x'} - \tilde w_i) \right) G_W \ .\nonumber
\ee  
Using that $\lim_{n \to 0} \sum_a (u^i_{ax} - \tilde w_i) = T U^i_x$ we obtain the
following decomposition 
\be 
\hat Z^{\phi^1 \phi^2}_{W}(\lambda_1,\lambda_2 ) = \hat Z^{\phi^1}(\lambda_1)  \hat Z^{\phi^2}(\lambda_2)  + \hat Z^{c,\phi^1 \phi^2}_{W}(\lambda_1,\lambda_2 )
\ee 
with the explicit forms
\bea \label{app:alg:disconn}
&& \hat Z^{\phi^1}(\lambda_1)  = \frac{ \int_x (-\nu_2 \psi^2_x + \int_{x'} g_{xx'}^{-1} U^{20}_{x'}) }{ \int_x  \psi^2_{x} }  \nn \\
&&  \hat Z^{\phi^2}(\lambda_2)  = \frac{ \int_x (-\nu_4 \psi^4_x + \int_{x'} g_{xx'}^{-1} U^{40}_{x'}) }{ \int_x  \psi^4_{x} } ~~~~~~~~~~
\eea
and
\bea \label{app:alg:conn}
 \hat Z^{c,\phi^1 \phi^2}_{W}(\lambda_1,\lambda_2 ) &=&  \frac{1}{\int   \psi^2_{x}  \int  \psi^4_{x} }  \int_{x} \int_{x'} g_{xx'}^{-1} U^{22}_{x'} \nn \\
&=&  \frac{1}{\int   \psi^2_{x}  \int   \psi^4_{x} }  \int_{x} \int_{x'} g_{xx'}^{-1} U^{41}_{x'}\ .~~~~~
\eea
Although not obvious,   these definitions are in agreement with those of the main text. Despite their complexity,     the equations satisfied by the $u$ and $U$ variables obey several symmetries. The important ones are $U^{10}_x = - U^{20}_x$ and $U^{30}_x = -U^{40}_x$; $U^{11}_x = -U^{21}_x$ and $U^{32}_x = -U^{42}_x$; $U^{12}_x = -U^{22}_x$ and $U^{31}_x = -U^{41}_x$;  $u^{11}_x = u^{21}_x$ and $u^{32}_x = u^{42}_x$;   $u^{12}_x = u^{22}_x$ and $u^{31}_x = u^{41}_x$. We also have $U^{22}_x = U^{41}_x$.

Using these symmetries, one finds that $U^{20}_x$ and $U^{40}_x$ satisfy 
\begin{eqnarray}
\int_{x'} g^{-1}_{xx'} U^{20}_{x'} = \sigma (U^{20}_x)^2 + \nu_2 \psi^2_x \ ,\nonumber \\
\int_{x'} g^{-1}_{xx'} U^{40}_{x'} = \sigma (U^{40}_x)^2+ \nu_4 \psi^4_x \ ,  
\end{eqnarray}
where $\sigma = R'''(0^+)$. Note that these are related to the function $Z_x^{\phi}(\lambda)$ defined in the main text in Eq.~(\ref{local:selfcons}) through the relation $Z_x^{\phi^1}(\lambda_1) = \int_{x'} g^{-1}_{xx'} U^{20}_{x'}$. Hence, Eq.\ (\ref{app:alg:disconn}) leading to the disconnected part of the result for $\hat Z^{\phi^1 \phi^2}_{W}(\lambda_1,\lambda_2 )$ is in agreement with the main text. Let us now introduce two important kernels defined as the functional derivatives $K_{2}(x,z) = \frac{ \delta U^{20}_x }{ \nu_2 \delta \psi^2_z } $ and $K_{4}(x,z) = \frac{ \delta U^{40}_x }{ \nu_4 \delta \psi^4_z } $. They satisfy
\bea \label{app:alg:eqnKernel}
&& \int_{x'} g_{xx'}^{-1} K_{2}(x',z) - 2 \sigma U^{20}_x K_2(x,z) = \delta(x-z) \nn \\
&& \int_{x'} g_{xx'}^{-1} K_{4}(x',z) - 2  \sigma U^{40}_x K_4(x,z) = \delta(x-z) \nn
\eea
and are important building blocks in our calculation. These kernels are symmetric: the kernel of the operator $K_2^{-1}$ is given by $K_2^{-1} (x , x') = g_{xx'}^{-1} - 2 \sigma U_{20x} \delta(x-x')$. In particular it is a symmetric function of its arguments, and thus $K_2(x,z)$  also is a symmetric function. The analytic expressions of the functions $U^{20}_{x}$ and $U^{40}_{x}$ are hard to obtain in generality. In Ref.\ 
\cite{LeDoussalWiese2008c} they were obtained for   avalanches measured on hyperplanes for SR elasticity: $\psi^2_x = \delta(x_1)$ where $x_1$ denotes the first coordinate of the $d$-dimensional variable $x$. We recall this explicit solution below in Appendix \ref{app:ZlocHyperplane}.

\subsubsection{Solutions for the $u$ variables}

Let us first consider the   solution for the $u$ variables. The equations read
\bea
&& \int_{x'} g_{xx'}^{-1} \Big(\frac{1}{2} - u^{11}_{x'}\Big) - 2 \sigma U^{10}_{x} u^{11}_{x}=0 \nn \\
&& \int_{x'} g_{xx'}^{-1} u^{31}_{x'} = 2 u^{11}_{x} U^{10}_{x} R'''(W) \nn \\
&& \int_{x'} g_{xx'}^{-1} \Big(\frac{1}{2} - u^{32}_{x'}\Big) - 2 \sigma U^{30}_{x} u^{32}_{x}=0 \nn \\
&& \int_{x'} g_{xx'}^{-1} u^{12}_{x'} = 2 u^{32}_{x} U^{30}_{x} R'''(W) 
\eea
The solutions are expressed in terms of the two kernels as
\bea
 u^{11}_{x} = u^{21}_{x} &=& - \frac{\sigma}{R'''(W)} u^{31}_{x} + \frac{1}{2} = - \frac{\sigma}{R'''(W)} u^{41}_{x} + \frac{1}{2}  \nn \\
  &=& \frac{m^2}{2} \int_{z} K_{2}(x,z)  \\
 u^{32}_{x} = u^{42}_{x} &=& - \frac{\sigma}{R'''(W)} u^{12}_{x} + \frac{1}{2} = - \frac{\sigma}{R'''(W)} u^{22}_{x} + \frac{1}{2}\nn \\
&  =& \frac{m^2}{2} \int_{z} K_{4}(x,z)
\eea

\subsubsection{Solutions for the $U$ variables}
For the $U$ variables, the equations read
\begin{align}
& \int_{x'}g_{xx'}^{-1}U^{21}_{x'} - 2 \sigma U^{20}_{x} U^{21}_{x}- 2 R^{(4)}(0) u^{11}_{x}(U^{20}_{x})^2=0 \nn \\
& \int_{x'}g_{xx'}^{-1}U^{42}_{x'} - 2\sigma U^{40}_{x} U^{42}_{x}- 2 R^{(4)}(0) u^{32}_{x}(U^{40}_{x})^2=0 \nn \\
& \int_{x'}g_{xx'}^{-1}U^{22}_{x'}  - 2\sigma U^{20}_{x} U^{22}_{x} -  2 R^{(4)}(W) u^{32}_{x}U^{20}_{x}U^{40}_{x}=0 \nn \\
& \int_{x'}g_{xx'}^{-1}U^{41}_{x'} - 2\sigma U^{40}_{x} U^{41}_{x}  - 2 R^{(4)}(W) u^{11}_{x}U^{40}_{x}U^{20}_{x}=0 \nn\\
\end{align}
Its solutions are
\bea
 U^{11}_{x} &=& -U^{21}_{x} = - 2 R^{(4)}(0) \int_z K_2(x,z) u^{11}_{z} (U^{20}_{z})^2 \nn\\
  U^{32}_{x} &=& -U^{42}_{x} = - 2 R^{(4)}(0) \int_z K_2(x,z) u^{32}_{z} (U^{40}_{z})^2
 \nn\\
 U^{12}_{x} &=& - U^{22}_{x} =- 2 R^{(4)}(W) \int_z K_2( x,z) u^{32}_{z} U^{20}_{z} U^{40}_{z} \nn\\
 U^{31}_{x} &=& - U^{41}_{x}  =- 2 R^{(4)}(W) \int_z K_4( x,z) u^{11}_{z} U^{20}_{z} U^{40}_{z} \nn\\
\eea

\subsubsection{Final result}
Using Eq.~(\ref{app:alg:conn}) we   obtain
\bea\label{app:localgene}
\lefteqn{ \hat{Z}_{W}^{c; \phi_1 , \phi_2} (\lambda_1 , \lambda_2)  =}  \\
&&  \frac{1}{ \int \psi_{x}^2  \int \psi_{x}^4 } R^{(4)}(W) m^4  \int_{ x' , z , z'} K_2( x',z) U^{20}_{z} U^{40}_{z}  K_{4}(z',z)  \nn 
\eea
Using  the above results $U^{20}_x = \int_{x'} g_{xx'} Z_{x'}^{\phi_1}(\lambda_1)$, and $U^{40}_x 
= \int_{x'} g_{xx'} Z_{x'}^{\phi_2}(\lambda_2)$, as well as  $K_2(x,z) = \int_{x'} g_{xx'} \frac{\delta Z_{x'}^{\phi^1}(\lambda_1) }{\lambda_1 \delta \phi^1_z}$ and  $K_4(x,z) = \int_{x'} g_{xx'} \frac{\delta Z_{x'}^{\phi^2}(\lambda_2) }{ \lambda_2 \delta \phi^2_z}$;  remembering that
$\psi^2_x=\phi^1_x$ and $\psi^4_x=\phi^2_x$,
one  shows that this formula is equivalent to Eq.~(\ref{localZc}).

\subsubsection{Simplified form of the final result}
The equivalent results (\ref{app:localgene}) and (\ref{localZc}) both involve  a functional derivative, which is in general a rather complicated object. We can however obtain a simplified formulation.  From Eq.~(\ref{app:localgene}) it is clear that it is sufficient to compute, for $i=1,2$,
\be  \label{app:alg:defpsi}
\chi_i(x) = \int_{z} K_i(z,x ) = \int_{z} K_i(x,z)
\ee 
rather than the full kernel $K_i$, and using the symmetry of $K_i$. Integrating Eq.~(\ref{app:alg:eqnKernel}) over $z$ one shows that $\chi_i(x)$ solves the equation
\bea \label{app:alg:eqnPsi}
&& \int_{x'} g_{xx'}^{-1}  \chi_2(x') - 2 \sigma U^{20}_x \chi_2(x) = 1 \ , \nn \\
&& \int_{x'} g_{xx'}^{-1} \chi_4(x') - 2  \sigma U^{40}_x \chi_4(x)  = 1 \ . \nn
\eea
Solving these equations (a task a priori simpler than the computation of the functional derivative) then leads to, following (\ref{app:localgene}),
\bea\label{app:localgene2}
\lefteqn{ \hat{Z}_{W}^{c;\phi_1 , \phi_2} (\lambda_1 , \lambda_2)  }  \\
&& = - \frac{1}{ \int \phi_{x}^1  \int \phi_{x}^2 } \Delta''(W) m^4  \int_{z} \chi_2(z) U^{20}_{z} U^{40}_{z}  \chi_4(z)  \ . \nn 
\eea

\subsection{More explicit solution for avalanches measured on parallel hyperplanes} \label{app:ZlocHyperplane}

\subsubsection{Setting}

We now obtain more explicit formulas in the case where avalanches are measured on two parallel hyperplanes at a      distance $y>0$ from one another and where the elasticity is short-ranged with kernel (\ref{short-rangedKernel}). That is, noting  for definiteness $x_1$ the first coordinate of the $d$-dimensional vector $x$, 
\be 
\phi^1_{x} = \delta(x_1) \quad, \quad \phi^2_{x} = \delta(x_1 - y) \ .
\ee 
In this case the problem becomes effectively unidimensional and the functions $U$ and $\chi$ entering into Eq.~(\ref{app:localgene2}) only depend on $x_1$, abbreviated as $x$ in the following. Furthermore, by translational invariance we can write
\bea
   U^{20}_{x} &=& Y(\lambda_1 , x) \quad , \quad \chi_2(x) = \chi(\lambda_1 , x)  \\
 U^{40}_{ x} &=& Y(\lambda_2 , x-y) \quad , \quad \chi_4(x) = \chi(\lambda_2 , x - y)   \ .\nn
\eea
These quantities obey the equations
\begin{align} \label{app:local:Y}
&   \left( - \frac{d^2}{dx^2} + m^2 \right) Y(\lambda ,x) - \sigma \left( Y(\lambda ,x) \right)^2 = \lambda \delta(x)   \\
& \left( - \frac{d^2}{dx^2} + m^2 \right) \chi(\lambda ,x) - 2 \sigma  Y(\lambda ,x)  \chi(\lambda ,x)  = 1 \nn .
\end{align} 
Solving these equations then leads to
\bea\label{app:localgene3}
\lefteqn{ \hat{Z}_{W}^{c;\phi_1 , \phi_2} (\lambda_1 , \lambda_2)  = \frac{1}{ L^{d-1} } R^{(4)}(W) m^4 \times} \\
&& \times   \int_{x}  \chi(\lambda_1 , x)  Y(\lambda_1 , x) Y(\lambda_2 , x - y) \chi(\lambda_2 , x- y)    \ . \nn 
\eea

\subsubsection{Solution for $Y$}

The solution $Y(\lambda, x)$ of  equation (\ref{app:local:Y}) is already known in the literature, see Ref.\ \cite{DelormeLeDoussalWiese2016} for details. It admits a scaling form
\be  \label{Yscaling}
Y(\lambda,x) = \frac{m^2}{\sigma} \tilde{Y}\Big( \frac{\sigma}{m^3} \lambda , mx\Big) \ , 
\ee 
where $\tilde{Y}(\tilde \lambda , \tilde x )$ solves
\be 
\left( - \frac{d^2}{d\tilde x^2} + 1 \right) \tilde Y(\tilde \lambda ,\tilde x) -  \left( \tilde Y(\tilde \lambda , \tilde x) \right)^2 = \tilde \lambda \delta(\tilde x) \ .
\ee 
An explicit solution is  
\be  \label{Yscaling2}
\tilde{Y}(\tilde \lambda , \tilde x ) = \frac{6(1 - z^2) e^{- | \tilde x| }  }{\left( 1 + z + (1-z) e^{- | \tilde x| }  \right)^2}\ ,
\ee 
where $z(\tilde \lambda)$ is one of the solutions of 
\be 
\tilde \lambda = 3 z (1 -z^2)  \ . 
\ee 
The right solution is uniquely defined from the following properties: it is defined for $\tilde \lambda \in ] - \infty  , \tilde \lambda_c = 2/\sqrt{3} [$, decreases from $z(-\infty) = \infty$ to $z_c = z (\tilde \lambda_c )= 1/\sqrt{3}$ and approaches $1$ as $\tilde \lambda$ approaches $0$.\\

\subsubsection{Solution for $\chi$}
From the coupled equations (\ref{app:local:Y}), it is   seen that $\chi(\lambda,x)$ can be deduced from $Y(\lambda,x)$ as
\be 
\chi(\lambda,x) =   \frac{1}{m^2} - \frac{2 \sigma}{m^2} \frac{\partial Y }{\partial m^2}  \  .
\ee 
Using the scaling form (\ref{Yscaling}) we obtain
\be \label{chiscaling1}
\chi(\lambda,x) = \frac{1}{m^2} \tilde \chi\left(\tilde \lambda= \lambda \frac{\sigma}{m^3},\tilde  x = mx\right)\ ,
\ee
where 
\be \label{chiscaling2}
\tilde \chi =  1 - 2 \tilde{Y} + 3 \tilde \lambda  \partial_{\tilde \lambda} \tilde Y -2 \tilde x 
\partial_{\tilde x}  \tilde Y    \  .
\ee

\subsubsection{Final scaling form}

Combining Eqs.\ (\ref{app:localgene3}), (\ref{Yscaling}) and (\ref{chiscaling1}) we can express our result in terms of a universal scaling function ${\cal Z}_{\tilde w}$ as (we scale $y =\tilde y/m$, $\lambda_i= \frac{m^3}{\sigma}\tilde{\lambda}_i$, $W = \tilde{w}/W_\mu$):
\bea
\lefteqn{ \hat{Z}_{\tilde{w}/W_{\mu}}^{c;\phi_1 , \phi_2} \left(\frac{m^3}{\sigma} \tilde{\lambda}_1 ,\frac{m^3}{\sigma} \tilde{\lambda}_2\right)} \nn\\
&=&   \frac{1}{(Lm)^{d-1} } \frac{1}{(mS_m)^2} \times  \hat {\cal Z}_{\tilde w}(\tilde \lambda_1 , \tilde \lambda_2 , \tilde y)  ~~~~~~
\eea
The quantities $W_\mu$ and $S_m$ are as in Eq.~(\ref{scalingWm}) with here $\mu = m$ (SR elasticity) and
\bea
\lefteqn{{\cal Z}_{\tilde w}(\tilde \lambda_1 , \tilde \lambda_2 , \tilde y) =  A_d \Delta^{* \prime \prime}(\tilde w)  \times} \\
&& \qquad \qquad \int_{\tilde x} \tilde \chi (\tilde \lambda_1 , \tilde x) \tilde Y (\tilde \lambda_1 , \tilde x)  \tilde Y (\tilde \lambda_2 , \tilde x - \tilde y )\tilde \chi (\tilde \lambda_2 , \tilde x - \tilde y )\ , \nn 
\eea 
where $\tilde Y$ and $\tilde \chi$ are explicit functions given in Eqs.~(\ref{Yscaling2}) and (\ref{chiscaling2}). This is our final result; its explicit evaluation is left for the future.

\section{First moment to one-loop order}

In this appendix we give the result for $\langle S_1^{2} S_2 \rangle_{\rho_W^c}$ to one-loop accuracy for short-ranged elasticity. Note that since the formula (\ref{S1S2exact}) is exact, it does not receive higher-loop contributions and the first improvement brought to moments of $\rho_W^c$ is for $\langle S_1^{2} S_2 \rangle_{\rho_W^c}$. The latter can be   obtained from the known   formulas (61) and (118) of Ref.~\cite{LeDoussalWiese2008c}, 
\bea
\lefteqn{\hat C^{(3)}(w_1 , w_2 , w_3) = -\frac{6}{m^2} {\rm  sym_{123}}  \Big\{\Delta'(w_{12}) \Delta(w_{13}) \Big\} }\nn\\
&& -6 I_3{\rm  \,sym_{123}} \Big\{ \Delta'(w_{12})^2  \Delta'(w_{13}) + [ \Delta(w_{12}) - \Delta(0)]    \nn \\
&&   ~~~~~~~~~~~~~~~~~\times \big[ \Delta'(w_{13} ) \Delta''(w_{12} ) + \Delta'(w_{12} ) \Delta''(w_{13} )  \nn\\ &&
~~~~~~~~~~~~~~~~~~~~~+ \Delta'(w_{23} ) \Delta''(w_{13} ) \big] \Big\}  
\ .\eea
The first line corresponds to the improved tree approximation, ${\rm  sym_{123}} $ denotes the symmetrization over the $w_i$ variables, $I_3 = \int_k \frac{1}{(k^2+m^2)^3}$, and we have use the shorthand notation $w_{ij} := w_i -w_j$. As explained in the text, this formula is sufficient to obtain $\langle S_1^{2} S_2 \rangle_{\rho_W^c}$ using the ${\cal K}$ operation. The final result reads
\bea
\lefteqn{ L^{-2 d} \langle S_1^2 S_2 \rangle_{\rho_W^c}  = - 6 S_m }\nn\\
&& - 4 I_3 \frac{S_m}{L^d m^2} \Big[ \Delta''(0) \Delta''(W) + 3 \Delta''(W)^2 \nn\\&& \qquad\qquad \qquad  + 3 \Delta'(W) \Delta'''(W)  \Big]  +O(\epsilon^3)  \ .\qquad 
\eea 

\bibliographystyle{ieeetr}
\bibliography{citation0915}

\tableofcontents

\end{document}